\documentclass[11pt,a4paper]{article}
\pdfoutput=1
\usepackage{jheppub_modified_green}
\usepackage{amsmath,amssymb,amsfonts,amsthm}
\usepackage{graphicx}
\usepackage{geometry}
\usepackage{xcolor}
\usepackage{pgfplots}
\usepackage{xparse}
\usepackage{hyperref}
\providecommand{\href}[2]{#2}
\usepackage{color}
\definecolor{light-gray}{gray}{0.9}
\newcommand{\bs}{\ensuremath{\kappa}}
\newcommand{\bc}{\ensuremath{\widetilde{\kappa}}}
\newcommand{\ds}[1]{
	\if 1#1
	\ensuremath{\left.\frac{\partial }{\partial \omega}\bs(t_1,-\omega)\right|_{\omega=0}}
	\else
	\ensuremath{\left.\frac{\partial^2 }{\partial \omega^2}\bs(t_1,-\omega)\right|_{\omega=0}}
	\fi
}
\newcommand{\dc}[1]{\if 1#1\ensuremath{\left.\frac{\partial }{\partial \omega}\bc(t_1,-\omega)\right|_{\omega=0}}\else\ensuremath{\left.\frac{\partial^2 }{\partial \omega^2}\bc(t_1,-\omega)\right|_{\omega=0}}\fi}

\makeatletter
\newcommand\makebig[2]{%
	\@xp\newcommand\@xp*\csname#1\endcsname{\bBigg@{#2}}%
	\@xp\newcommand\@xp*\csname#1l\endcsname{\@xp\mathopen\csname#1\endcsname}%
	\@xp\newcommand\@xp*\csname#1r\endcsname{\@xp\mathclose\csname#1\endcsname}%
}
\makeatother

\makebig{biggg} {3.0}
\makebig{Biggg} {3.5}
\makebig{bigggg}{4.0}
\makebig{Bigggg}{4.5}

\usetikzlibrary{positioning,arrows,patterns}
\usetikzlibrary{decorations.markings}
\usetikzlibrary{calc}

\tikzset{
	boson/.style={draw=black},
	photon/.style={decorate, decoration={snake}, draw=black},
	fermion/.style={draw=black, postaction={decorate},decoration={markings,mark=at position .55 with {\arrow{>}}}},
	vertex/.style={draw,shape=circle,fill=black,minimum size=3pt,inner sep=0pt},
}

\NewDocumentCommand\semiloop{O{black}mmmO{}O{above}}
{%
\draw[#1] let \p1 = ($(#3)-(#2)$) in (#3) arc (#4:({#4+180}):({0.5*veclen(\x1,\y1)})node[midway, #6] {#5};)
}

\title{Dynamics of scalar fields in an expanding/contracting cosmos at finite temperature}
\author[a]{Hui Xu,}
\author[a]{Lei Ming,}
\author[a,b]{Yeuk-Kwan E. Cheung,}

\affiliation[a]{School of Physics, Nanjing University, Nanjing, 210093, China}
\affiliation[b]{Institute of Nuclear and Particle Physics,
Demokritos National Research Centre,  Athens,  Greece}
\emailAdd{minglei@smail.nju.edu.cn}
\emailAdd{huixu@smail.nju.edu.cn}
\emailAdd{cheung@nju.edu.cn}

\abstract{{
	This paper extends the study of the quantum dissipative effects of a cosmological scalar field  by taking into account  the  cosmic  expansion  and  contraction.  
	Cheung, Drewes, Kang and  Kim~\cite{Cheung:2015iqa}  calculated the effective action and quantum dissipative effects of a cosmological scalar field.  
	The analytic expressions for the effective potential and damping coefficient were presented using a simple scalar model with quartic interaction. 
		Their work was done using Minkowski-space propagators in loop diagrams.   
	In this  work we incorporate  the Hubble  expansion and contraction  of the comic background,  and focus on  the thermal dynamics  of a scalar field in a regime where the effective potential changes slowly. 
	We let the Hubble parameter, $H$,  attain  a small but non-zero value and carry out  calculations to first order in $H$. 
	If we set $H=0$  all results match those~\cite{Cheung:2015iqa}  in flat spacetime.  
	Interestingly   we have to integrate  over the resonances, which in turn leads to an amplification of the effects of a non-zero $H$. 
	This is an intriguing phenomenon  which cannot be uncovered in  flat spacetime. 
	The implications on particle creations  in  the early universe will be studied in a forthcoming work. 
}}


\begin{document}
	\maketitle
	\flushbottom
	\newpage
	\raggedbottom
	
\section{Introduction}

It is widely believed that our universe starts  with a hot big bang,  which is considered as the beginning of the radiation dominated  era  in the cosmic history. Many properties of the cosmos that we observe today can be understood as the results of quantum processes, which would be typically out of equilibrium, in the hot and dense plasma~\cite{Gamow:1946eb, Kolb:1990vq} that filled the universe after the big bang.  Prior to the radiation dominated era, there was a period of accelerated cosmic expansion known as inflation~\cite{Starobinsky:1980te, Guth:1980zm, Linde:1981mu}. 
At  the end of inflation  the universe is  cold and empty;  all energy is stored in the zero mode of the inflaton field. 
One mechanism for setting up the ``hot big bang''  initial conditions of a radiation dominated universe is ``reheating''~\cite{Kofman:1994rk, Shtanov:1994ce,Drewes:2013iaa}, in which 
the universe is  ``reheated"  from a complete vacuum by the energy transfer from the inflaton to other degrees of freedom, e.g.  dark matter particles and  elementary particles that made up  the Standard Model of Particle Physics.   
The study of the quantum dissipative effects in the early universe, therefore, has profound implications on the studies of matter production and thus on the thermal history of our universe. 

The thermal history of the early universe is an important theoretical basis to determine the abundance of thermal relics.  It thus plays an important role in distinguishing  or excluding  cosmological models.   
Studies of the particle dynamics in the  early universe uncover  crucial details  within and beyond the Standard Model.  
We are interested in the  thermal production of particles from plasma~\cite{Berera:2004kc, Graham:2008vu},  dissipation effects  of fields in medium~\cite{Lee:1999iv, deOliveira:1997jt}, cosmological freeze-out processes~\cite{Joyce:1996cp, Joyce:1997fc} and their imprints  on  early universe physics.  

As we have mentioned above,  matter production is via  the relaxation of inflaton into scalar, gauge and fermionic quantum fields in a large thermal bath~\cite{Berera:2004kc, Yokoyama:2004pf, Anisimov:2008dz, Bartrum:2014fla}.  
Inflaton, in the standard model of cosmology,  is a scalar and responsible for an epoch of exponential expansion to produce a flat, homogeneous and isotropic universe free of topologically stable relics like monopoles and cosmic strings. 
Therefore, scalar fields, despite being the simplest, play important roles within or beyond the Standard Models of Particle Physics and Modern Cosmology. 
The existence of scalar fields in the Standard Model of Particle physics  has been firmly established by  high precision experiments conducted  at the Large Hadron Collider in 2012, where  the  Higgs boson~\cite{Aad:2012tfa, Chatrchyan:2012xdj} plays a pivotal role of giving mass to elementary particles in the standard model.  
In addition, scalar fields can be candidates for dark energy~\cite{Wetterich:1987fm, ArmendarizPicon:2000dh, Copeland:2006wr} or dark matter~\cite{McDonald:1993ex, Burgess:2000yq, Bento:2000ah}.  
Scalar fields also play important roles in the bounce universe which is an alternative way to address how our current  universe comes about.  In this model  a contraction phase exists  prior to the ``birth''  of our presently observed universe, see~\cite{Brandenberger:2016vhg, Battefeld:2014uga} for recent reviews. 
Our study focuses on the scalar field dynamics in hot medium:  it thus founds the basis to follow up on the particle productions in these  bounce models~\cite{Cai:2011zx, Lin:2010pf, Allen:2004vz, Khoury:2001wf, Cai:2013kja, Loewenfeld:2009aw,  Li:2011nj},  where the Hubble parameter can be set  to   positive or negative values,  and the bounce process is  driven  by  two or more scalar fields. 

This paper  builds  on  an  earlier study of  the finite-temperature effects in a thermal bath, carried out by Y. K. E. Cheung, M. Drewes, J. U Kang and J. C. Kim~\cite{Cheung:2015iqa},  to further establish the  rigorous theoretical framework to  precisely study  the evolution and  interactions  of elementary fields  in the inflationary cosmology background or in a bounce universe. 
In~\cite{Cheung:2015iqa}, the authors have made progress towards a quantitative understanding of the  non-equilibrium dynamics of scalar fields in the non-trivial background of the early universe with a high temperature,  large energy density and a rapid cosmic expansion. They calculated the effective action and quantum dissipative effects of a cosmological scalar field in this background.  The analytic expressions for the effective potential and damping coefficient are presented using a simple scalar model with quartic interaction.  In this paper, we extend their efforts on building this theoretical framework  by incorporating a non-zero Hubble parameter in the analysis and obtain temperature dependent expression of the damping coefficient to first order in $H$.  In this way one can properly  address the questions of how the hot primordial plasma may have been created after inflation~\cite{Drewes:2010pf} and whether there are observable features of the reheating process~\cite{Drewes:2015coa, Drewes:2017fmn}.  

Our study of the non-equilibrium process in an early universe starts from the action of scalar fields  $\phi$ and $ \chi $. The non-equilibrium process is non-Markovian. That is, the evolution of the fields in late times depends on all the previous states, and the non-Markovian effects are contained in a ``memory integral'' in the Kadanoff-Baym equations. 
We shall use the closed-time-path (CTP) formalism~\cite{Schwinger:1960qe,Keldysh:1964ud},  the so-called  ``in-in formalism.''  The  ``in-in formalism"  is made to deal with such finite temperature problems in out-of-thermal-equilibrium processes.  The non-equilibrium nature renders the usual  ``in-out formalism''  ineffective. 

Unlike the usual zero-temperature quantum field theory, this paper considers both the thermal corrections and  quantum corrections. To be more specific, we first derive the equation of motion, up to the first order in $ H $,  from the effective action of $\phi$.  The effective potential and the dissipation coefficient (characterising the dissipation of the energy from 
 $\phi$ to the plasma)  rely on the self-energy and the corrections to the quartic coupling constant.  
The calculations of  these  quantities  using Feynman diagrams  make up the major part of work  reported in this article. 
  If we set $H$ to 0,  our  results  match up with  those obtained, in~\cite{Cheung:2015iqa},  using a Minkowski-space propagator in  loops.   In addition we observe non-trivial features that are not revealed in flat spacetime.

Exposing scalar fields to a high temperature and a rapid cosmic expansion is an important setup for understanding the non-equilibrium dynamics of scalar fields in cosmology.  
Under the condition of cosmic expansion at high temperature, the matter production process is out of equilibrium.  If there is an effective potential that is not steep, the matter fields take a long time to reach equilibrium.  Our focus is the back reaction~\cite{Vachaspati:2018llo, Calzetta:1986ey} on the primary particles from  their decay products. 
Based on the previous work done using Minkowski-space propagators in loop diagrams,  we further their studies of scalar field dynamics  in the early universe evolution by incorporating the effect of cosmic expansion.  There are infinitely many back reactions, and thus it is important to generalise the leading-order re-summation results to higher orders.

Although elementary  particles in our universe consist of  fermions and gauge  bosons,  we still expect our toy model with two  scalar fields  to  serve as a good playground for studying  the early universe physics.  
The earliest decay process involves only scalars,  because  the creation of fermions can be assumed to happen in the subsequent decay  chain or inelastic scatterings as  the transition to bosonic states is usually Bose-enhanced~\cite{Drewes:2013iaa}.  When the background temperature is higher than the oscillation  frequency,  the dissipation rate arising from the interactions with fermions is suppressed due to Pauli blocking,  whereas  it is enhanced for interactions with bosons due to the induced effect.  In a future work, we will consider the direct coupling of scalars to fermions and gauge bosons.

This paper is organised  as follows: In section~\ref{sec:assumptions}, we explain  our  working assumptions.  We sketch  the prerequisites for field theory calculations at finite temperature to establish notations,  which is  followed by the derivation of the EoM in section~\ref{sec:EoM}. We demonstrate the calculations of self-energy and  obtain the  correction to the quartic coupling constant in section~\ref{sec:computation}. Section~\ref{sec:discussion} briefly summarises   the main results obtained in this paper and concludes  with a short discussion.  
The detailed calculations of  the  Feynman diagrams and the relevant  formulae needed in  the calculations are contained in  the  appendices.
	
\section{Assumptions and Prerequisites}
\label{sec:assumptions}

We study a scalar field, denoted by $\phi$,  in a de Sitter space with the Hubble parameter, $H$, taken to be nonzero.  The scalar field,  $\phi$,  interacts with another scalar field, $ \chi $,  which plays the role of the cosmic background in our model.  The mass of the thermal bath, $\chi$,  is assumed to be less than half the mass of $\phi $.
$\phi$ can be decomposed into its thermal average, $ \varphi $,  and  fluctuations,  denoted by $ \eta $: 
$ \phi=\varphi+\eta $.
The fields,  $\eta $ and $ \chi $,   are assumed to be in thermal equilibrium initially.
We assume the reaction between $ \varphi $ and other degrees of freedom is weak enough to allow for the application of  perturbation theory and  the  neglecting  of  back reaction of $\varphi$ on  the $ \eta $ and $ \chi $ fields. 
In this way   $\eta$ and $\chi$ can  assume thermal equilibrium  in the evolution of $ \varphi $.
In order that we can expand the equation of motion to  first order in $ H $ and simplify the calculations,  we further assume $ H < m_\phi, m_\chi\ll T $($ m_\phi $ and $ m_\chi $ are the masses of the $ \phi $ field and the $ \chi $ field, respectively) and $ T $ is inversely proportional to the scale factor as the universe expands. 
There exists  a prolonged period of time in our universe when these conditions are satisfied.

\begin{figure}[h]
	\begin{center}
		\includegraphics[width=10cm]{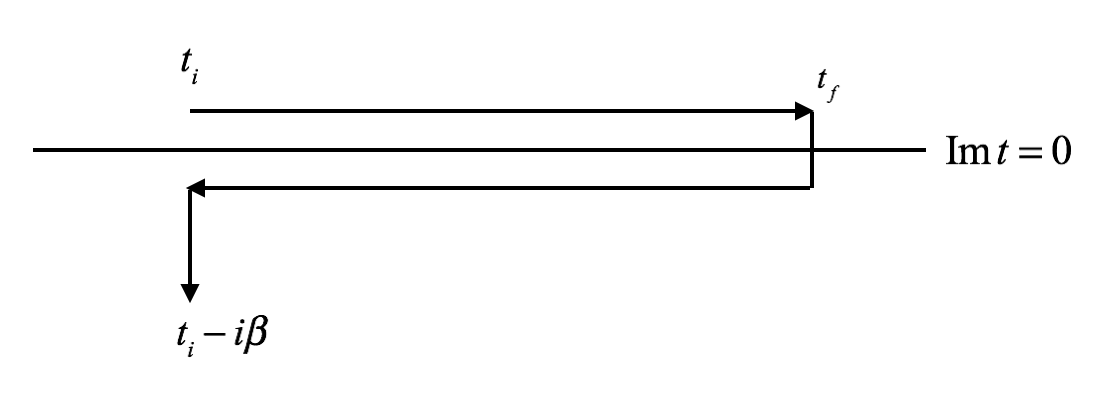}
	\end{center}
	\caption{The path of integration in the complex-time plane}
	\label{fig:path}
\end{figure}

We use the closed-time-path formalism to perform the calculations.
The time ordering and integration path $C$ in the complex time plane starts from $ t_i $ on the real axis and runs to real $t_f$, then back to $ t_i $ and ends at $ t_i-i\beta $, as depicted in Fig.~\ref{fig:path}.  
We take $ t_i \to -\infty$ and $ t_f\to\infty $ and denote the upper section of $ C $ which runs forward in time by $ C_1 $ and the lower section which runs backward in time by $ C_2 $.
And a general scalar field $ \xi(x) $ that lies on $ C_1/C_2 $ is labelled as, respectively, $ \xi_1(x) $/$ \xi_2(x) $ .

In order to obtain the dynamical information of such a scalar field $ \xi $, we need to know the two-point correlation functions which are defined as follows,
\begin{eqnarray*}
	\begin{array}{rclrcl}%
		\Delta_{ab}(x,y)&=&\langle T_C\xi_a(x)\xi_b(y) \rangle & \quad  (a,b &=&1,2),\\
		\Delta^>(x,y) &=& \langle\xi(x)\xi(y)\rangle,   &\quad  \Delta^<(x,y)  &=&\langle\xi(y)\xi(x)\rangle,\\
		\Delta^-(x,y) &=& i\left[\Delta^>(x,y)-\Delta^<(x,y)\right],  
		        &\quad \Delta^+(x,y) &=&\frac{1}{2}\left[\Delta^>(x,y)+\Delta^<(x,y)\right],
	\end{array}
	\end{eqnarray*}
where $ T_C $ indicates the  time ordering along the path $ C $ in Fig.~\ref{fig:path}.
For a real scalar field $ \xi $, we see that $ \Delta^{>}(x,y)={(\Delta^<(x,y))}^{*} $ and that
\begin{equation}\label{propagator_identity}
	\begin{aligned}
		&\Delta_{11}(x,y)=\theta(x^0-y^0)\Delta^{>}(x,y)+\theta(y^0-x^0)\Delta^{<}(x,y),\\
		&\Delta_{22}(x,y)=\theta(x^0-y^0)\Delta^{<}(x,y)+\theta(y^0-x^0)\Delta^{>}(x,y),\\
		&\Delta_{12}(x,y)=\Delta^{<}(x,y),\\
		&\Delta_{21}(x,y)=\Delta^{>}(x,y).
	\end{aligned}
\end{equation}

The 1-loop propagators of a scalar field $ \xi $ with effective mass $ M_\xi $ in a de-Sitter space, 
\begin{equation}\label{metric}
g_{\mu\nu}=\text{diag}(1,-a(t)^2,-a(t)^2,-a(t)^2),  
\end{equation}
where  $a(t)= e^{Ht}$, 
were obtained  by solving the Kadanoff-Baym equations~\cite{Drewes:2012qw}. 
The result in~\cite{Drewes:2012qw} is expressed in terms of conformal time, with a time dependent mass $ m(t) $. 
To obtain the corresponding result in terms of the cosmic time, we first carry out the usual replacement: 
\begin{equation}
t\to -\frac{1}{Ha(t)}~.
\end{equation}
The  next  two replacements  can be inferred by comparing the free spectral function in  de Sitter spacetime 
(the   detailed calculation of which is presented in Appendix~\ref{appendix:free-spectral-function})
with  the corresponding  flat-spacetime propagator~\cite{Drewes:2012qw}:
\begin{equation}
\xi(x)\to \xi(x)/a(t),\quad m(t)^2\to \left(M_\xi^2-2H^2\right)a(t)^2~.
\end{equation}

We thus start  our calculations with  the following  propagators in de Sitter spacetime:  
\begin{equation} 
\left\{
\begin{aligned}
&\Delta^{-}(\boldsymbol{p},t_1,t_2)=\dfrac{\displaystyle\sin\left(\int_{t_2}^{t_1}dt'\Omega_\xi(t')\right)\exp\left(-\frac{1}{2}\left|\int_{t_2}^{t_1}dt' \Gamma_\xi(t')\right|\right)}{a(t_1)^{3/2}a(t_2)^{3/2}\sqrt{\Omega_\xi(t_1)\Omega_\xi(t_2)}}\\
&\Delta^{+}(\boldsymbol{p},t_1,t_2)= \left[1+2f(t_B)\right] \cdot \dfrac{\displaystyle\cos\left(\int_{t_2}^{t_1}dt'\Omega_\xi(t')\right)\exp\left(-\frac{1}{2}\left|\int_{t_2}^{t_1}dt' \Gamma_\xi(t')\right|\right)}{2a(t_1)^{3/2}a(t_2)^{3/2}\sqrt{\Omega_\xi(t_1)\Omega_\xi(t_2)}}~,
\end{aligned}
\right.
\end{equation}
where $ \Gamma_\xi $ is the decay rate of $ \xi $, $\Omega_\xi$ is given by 
 \begin{equation}
 	\Omega_\xi(t)=\sqrt{\boldsymbol{p}^2/a(t)^2+M_\xi^2-2H^2}~,
 \end{equation}
with $ t_B= \min(t_1,t_2) $,  and 
$ f(t) $ satisfies a  Markovian equation~\cite{Drewes:2012qw}. 

If we assume that a scalar field $ \xi $ is in (approximate) thermal equilibrium as the universe expands,  and  can be characterized by an effective temperature $ 1/\beta(t) $ which depends on time,  then by imposing the KMS relations, the propagators become
\begin{equation}\label{propagator+-}
\left\{
\begin{aligned}
\Delta^{-}(\boldsymbol{p},t_1,t_2)=&\dfrac{\displaystyle\sin\left(\int_{t_2}^{t_1}dt'\Omega_\xi(t')\right)\exp\left(-\frac{1}{2}\left|\int_{t_2}^{t_1}dt' \Gamma_\xi(t')\right|\right)}{a(t_1)^{3/2}a(t_2)^{3/2}\sqrt{\Omega_\xi(t_1)\Omega_\xi(t_2)}}\\
\Delta^{+}(\boldsymbol{p},t_1,t_2)=&\dfrac{1}{2a(t_1)^{3/2}a(t_2)^{3/2}\sqrt{\Omega_\xi(t_1)\Omega_\xi(t_2)}}\exp\left(-\frac{1}{2}\left|\int_{t_2}^{t_1}dt' \Gamma_\xi(t')\right|\right)\\
&\times\left\{\displaystyle\exp\left(-i\int_{t_2}^{t_1}dt'\Omega_\xi(t')\right)\left[\frac{1}{2}+f\left(\Omega_\xi\left(t_B\right)+\frac{i\Gamma_\xi\left(t_B\right)}{2}\right)\right]\right.\\
&\left.+\exp\left(i\int_{t_2}^{t_1}dt'\Omega_\xi(t')\right)\left[\frac{1}{2}+f\left(\Omega_\xi\left(t_B\right)-\frac{i\Gamma_\xi\left(t_B\right)}{2}\right)\right]\right\}~,
\end{aligned}
\right.
\end{equation}
where $ f $ becomes the Bose distribution function:
\begin{equation}
f(x)=\frac{1}{e^{\beta(t)x}-1}~.
\end{equation}
Here the inverse temperature $ \beta(t) $ is proportional to the scale factor by our assumptions: 
$ \beta(t)=(\beta_0/a_0)a(t) $ with $ \beta_0 $ being the inverse temperature when the scale factor is $ a_0 $. For simplicity, we will denote $ \beta_0/a_0 $ by $ \gamma $ in the following calculations.
For more details about field theory at finite temperature,  the readers are referred to~\cite{Bellac:2011kqa, Drewes:2012qw}.

\section{Derivation of the EoM} \label{sec:EoM}
As we have mentioned in Introduction,   our system consists of a  scalar field $ \phi $ and  the background plasma collectively denoted by $ \chi $. 
We use the model to  study the early universe dynamics: it gives us clues on how the fields behave in an expanding or contracting universe.  In particular we wish to capture  how their behaviour differs as  one goes beyond using the Minkowski propagator in computing the quantum and thermal corrections. 
A general renormalizable action for such a system in a de-Sitter space,  whose potential energy is bounded from below, is of the form:  
\begin{equation}\label{eqn:action}
	\begin{aligned}
		S=&-\int d^4x\sqrt{-g}\left\{\frac{1}{2}\phi\left[\frac{1}{\sqrt{-g}}\partial_{\mu}\left(\sqrt{-g}g^{\mu\nu}\partial_{\nu}\right)+m_\phi^{2}\right]\phi+\frac{\lambda}{4!}\phi^4\right.\\
		&\left.+\frac{1}{2}\chi\left[\frac{1}{\sqrt{-g}}\partial_{\mu}\left(\sqrt{-g}g^{\mu\nu}\partial_{\nu}\right)+m_\chi^{2}\right]\chi+\frac{\lambda'}{4!}\chi^4+\frac{h}{4}\phi^2\chi^2\right\}.
	\end{aligned}
\end{equation}
The action is furthermore invariant under  diffeomorphism and  $ \phi(x)\to -\phi(x) $.  

The effective action for $ \varphi $ (which can be used to obtain the equation of motion of $ \varphi $)  has the same linear symmetries as the original action~\cite{Weinberg:1996kr},  it can be written in the form,
\begin{equation}\label{effective_action}
	\begin{aligned}
		\Gamma&=-\frac{1}{2}\int_C d^4x_1\sqrt{-g}\varphi(x_1)\left[\frac{1}{\sqrt{-g}}\partial_{1\mu}\left(\sqrt{-g}g^{\mu\nu}\partial_{1\nu}\right)+m_\phi^{2}\right]\varphi(x_1)\\
		&\hphantom{=}-\frac{1}{2}\int_C \sqrt{-g(x_1)}\sqrt{-g(x_2)}d^4x_1d^4x_2\Pi(x_1,x_2)\varphi(x_1)\varphi(x_2)\\
		&\hphantom{=}-\int_C\sqrt{-g(x_1)}\sqrt{-g(x_2)}d^4x_1d^4x_2\frac{1}{4!}\left[\lambda\frac{\delta^4(x_1-x_2)}{\sqrt{-g(x_1)}}+\widetilde{\Pi}(x_1,x_2)\right]\varphi(x_1)^2\varphi(x_2)^2,
	\end{aligned}
\end{equation}
up to the fourth power of $ \varphi $. 
In doing so we  have only considered terms up to 1-loop in the quartic part, and the time integration is along the path  $C$ shown in Fig.~\ref{fig:path}.  $ \Pi $ is the self-energy,  and $ \widetilde{\Pi} $  the correction to the quartic coupling constant,  the computation of  which will be presented  in the next section.  
Similar to the case of propagators,  we will denote $ \Pi(x_1,x_2) $ as $ \Pi_{ab}(x_1,x_2) $ when $ x_1^0 $ lies on  $C_a$ and $x_2^0$ lies on $ C_b $ ($a,b=1\, \text{or}\, 2$).   
Different from the situation in flat spacetime, here we do not have time translation symmetry:  $ \Pi(x_1,x_2),\widetilde{\Pi}(x_1,x_2) $ depend not only on $ t_1-t_2 $, but on $ t_1+t_2 $ as well.

From~(eq.~\ref{effective_action}) we obtain the equation of motion for $ \varphi $,
\begin{gather}
	\begin{aligned}
		\frac{\delta\Gamma}{\delta\varphi(x_1)}&=e^{3Ht_1}\left(\partial_{0}^{2}+3H\partial_0-e^{-2Ht_1}\nabla^2+m_\phi^2\right)\varphi(x_1)+\int_C dx_2 e^{3H(t_1+t_2)}\Pi(x_1,x_2)\varphi(x_2)\\
		&\hphantom{=}+\frac{\lambda}{3!}\varphi(x_1)^3e^{3Ht_1}+\frac{1}{3!}\varphi(x_1)\int_C dx_2e^{3H(t_1+t_2)}\widetilde{\Pi}(x_1,x_2)\varphi(x_2)^2\,=\, 0~. 
	\end{aligned}
\end{gather}
We restrict ourselves to the case in which  the only non-vanishing Fourier mode of $ \varphi(x) $ is $ \varphi(\boldsymbol{q}=0) $, then in (spatial) momentum space, the EoM simplifies to
\begin{gather}\label{zeroModeEoM}
	\begin{aligned}
		\left(\partial_{0}^{2}+3H\partial_0+m_\phi^2\right)\varphi(t_1)+\int_C dt_2 e^{3Ht_2}\Pi(t_1,t_2)\varphi(t_2)&\\
		\hphantom{=}+\frac{\lambda}{3!}\varphi(t_1)^3+\frac{1}{3!}\varphi(t_1)\int_C dt_2e^{3Ht_2}\widetilde{\Pi}(t_1,t_2)\varphi(t_2)^2 &=0~.
	\end{aligned}
\end{gather}
This non-local equation is still hard to solve.  
To make progress the Hubble parameter $ H $ is assumed to be  small and the system is in pseudo-equilibrium during the process.  This is a valid hypothesis for the most of the physical applications we have in mind. 
We  can hence simplify the equation by expanding $e^{3Ht_2} \approx1+3Ht_2 $ to first order in $ H$;
while the  adiabatic assumption is realised as,  $ \varphi(t_2)\approx \varphi(t_1)+\dot{\varphi}(t_1)(t_2-t_1)$.  
Altogether, 
\begin{equation}
	\begin{aligned}
		\left(\partial _{0}^{2}+3H\partial _0+m_\phi^2\right)\varphi(t_1)+\int_{-\infty}^{\infty}dt_2(1+3Ht_2)\Pi^R(t_1,t_2)\left[\varphi(t_1)+\dot{\varphi}(t_1)(t_2-t_1)\right]&\\
		+\frac{\lambda}{3!}\varphi(t_1)^3 +\frac{1}{3!}\varphi(t_1)\int_{-\infty}^{\infty}dt_2(1+3Ht_2)\widetilde{\Pi}^R(t_1,t_2)\left[\varphi(t_1)^2+2\varphi(t_1)\dot{\varphi}(t_1)(t_2-t_1)\right] &=0 ~. 
	\end{aligned}
\end{equation}
Let $\Pi^{R}(t_1,t_2) $ denote the retarded propagator: 
\begin{equation}
	 \Pi^{R}(t_1,t_2)=\Pi_{11}(t_1,t_2)-\Pi_{12}(t_1,t_2),
\end{equation}
and likewise, 
\begin{equation}
	\widetilde{\Pi}^R(t_1,t_2)=\widetilde{\Pi}_{11}(t_1,t_2)-\widetilde{\Pi}_{12}(t_1,t_2);
\end{equation}
and the Fourier transformation  of $\Pi^R(t_1,t_2)$ and $ \widetilde{\Pi}^R(t_1,t_2) $ (with respect to $t=t_1-t_2$)
be denoted by   $\kappa(t_1,\omega) $ and $\widetilde{\kappa}(t_1,\omega) $: 
\begin{equation}
	\begin{aligned}
		&\kappa(t_1,\omega)\equiv \int dt\,\,\Pi^R(t_1,t_1-t)e^{i\omega t},\\
		&\widetilde{\kappa}(t_1,\omega)\equiv \int dt\,\,\widetilde{\Pi}^R(t_1,t_1-t)e^{i\omega t}~.
	\end{aligned}
\end{equation}
The equation of motion then  becomes, 
\begin{equation}\label{eqn: effective_EoM}
	\begin{aligned}
		&\ddot{\varphi}(t_1)+\Bigg\{\left[m_\phi^2+(1+3Ht_1)\bs(t_1,\omega=0)-3H\left(-i\ds{1}\right)\right]\varphi(t_1)\\
		&\hphantom{\ddot{\varphi}(t_1)+}
		+\frac{1}{6}\left[\lambda+(1+3Ht_1)\bc(t_1,\omega=0)-3H\left(-i\dc{1}\right)\right]\varphi(t_1)^3\Bigg\}\\
		&+\Bigg\{3H+(1+3Ht_1)\left(-i\ds{1}\right)+3H\ds{2}\\
		&\hphantom{+}
		+\frac{1}{3}\left[(1+3Ht_1)\left(-i\dc{1}\right)+3H\dc{2}\right]\varphi(t_1)^2\Bigg\}\,\dot{\varphi}(t_1) =0~.
	\end{aligned}
\end{equation}
From this equation of motion we see that the ``potential'' (terms in the first curly bracket)
which is time-dependent for $ \varphi $,  is determined by 
$$\bs(t_1,\omega=0), ~~\bc(t_1,\omega=0), ~~\ds{1}, ~~\dc{1} ; $$    
whereas  the dissipation coefficient (terms in the second curly bracket) relies on  
$$\ds{1}, ~~\dc{1}, ~~\ds{2}, ~~\dc{2} . $$
We will obtain  these expressions in the following section.

\section{Computation of the potential and the dissipation coefficient}
\label{sec:computation}

In this section  we shall  calculate the self-energy and the correction to the self-coupling constant of $\phi$,  together with  the first and second derivatives of them.

\subsection{Computation of $\bs(t_1)$}   \label{sec:kappa}
\begin{figure}
	\centering
	\begin{tikzpicture}[scale=2]
	\draw (-1,0) -- (0,0) node [align=center, below]{$ t_1=t_2 $} -- (1,0);
	\draw (0,0.5) circle(0.5);
	\end{tikzpicture}
	\caption{A Feynman  diagram, the tadpole diagram,  corresponding to the real part of $ \Pi^R. $}
	\label{fig:tadpole}
\end{figure}
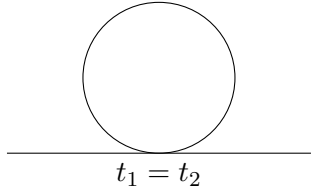

The leading order contribution to $ \Pi^R(t_1,t_2) $,  and $ \kappa(t_1) \equiv \kappa(t_1,\omega=0)$,  is  given by the tadpole diagram, shown in Fig.~\ref{fig:tadpole},  with $ \eta $ or $ \chi $ running in the loop. 
We observe that  $\Pi_{12}=0 $  because $ t_1 $ and $ t_2 $ must be identical.  
Using~(eq.~\ref{propagator_identity})  we obtain,
\begin{equation}\label{eqn:tadpole}
	\begin{aligned}
		 \Pi^{R}(t_1,t_2)&=\Pi_{11}(t_1,t_2)\\
		&=\frac{1}{2}\delta(t_1-t_2)\sum_{\xi=\eta,\chi}\left[g_{\xi}\int \frac{d^3p}{(2\pi)^3}(\Delta_\xi)_{11}(\boldsymbol{p},t_1,t_2)\right]\\
		&=\frac{1}{2}\delta(t_1-t_2)\sum_{\xi=\eta,\chi}\left[g_{\xi}\int \frac{d^3p}{(2\pi)^3}\mathrm{Re}(\Delta_\xi)_>(\boldsymbol{p},t_1,t_2)\right]~, 
	\end{aligned}
\end{equation}
with  $ g_\eta\equiv\lambda, g_\chi\equiv h $. 
The integral  is  computed  in Appendix~\ref{appendix: bose distribution integral},
\begin{equation} \label{eqn:Pit1t2}
	\begin{aligned}
	\Pi^R(t_1,t_2)&=\delta(t_1-t_2)\left\{\frac{1}{2\pi^2\beta(t_1)^2}\sum_{\xi=\eta,\chi}g_{\xi}h_3    	\left[M_\xi\beta(t_1)\right]\right\}~. 
	\end{aligned}
\end{equation}
$ M_\eta$ and $M_\chi $ denote,  respectively, the effective masses for $\eta$ and $\chi$,  and $ h_3 $ is given by, 
\begin{equation}   \label{eqn: h_3}
h_3(y)
=\frac{\pi^2}{12}-\frac{\pi y}{4}-\left(\frac{\gamma_E}{8}-\frac{1}{16}\right)y^2-\frac{y^2}{8}\log\frac{y}{4\pi}
+\sum_{m=1}^{\infty}\frac{(-1)^{m+1}}{2^{m+3}}\frac{(2m-1)!!}{(m+1)!}\frac{\zeta(2m+1)}{(2\pi)^{2m}}y^{2m+2}~, 
\end{equation}
with $ \gamma_E $ being the Euler constant.  Fourier transforming~(eq.~\ref{eqn:Pit1t2}) yields,
\begin{equation}\label{eqn:kappa}
	\begin{aligned}
		\bs(t_1)&=\frac{1}{2\pi^2\beta(t_1)^2}\sum_{\xi=\eta,\chi}g_{\xi}h_3\left[M_\xi\beta(t_1)\right]\\
		&\approx \frac { \lambda + h } { 24 \gamma ^ { 2 } } \left( 1 - 2 H t _ { 1 } \right).
	\end{aligned}
\end{equation}

\subsection{Computation of $\bc(t_1) $}
\begin{figure}
	\centering
	\begin{tikzpicture}
	\coordinate[vertex,label=left:$ t_1 $](v1);
	\coordinate[vertex,right=of v1,label=right:$ t_2 $](v2);
	\coordinate[vertex,above left=of v1](e1);
	\coordinate[vertex,below left=of v1](e2);
	\coordinate[vertex,above right=of v2](e3);
	\coordinate[vertex, below right=of v2](e4);
	\draw[boson](e1)--(v1);
	\draw[boson](v1)--(e2);
	\draw[boson](v2)--(e3);
	\draw[boson](v2)--(e4);
	\semiloop[boson]{v1}{v2}{0};
	\semiloop[boson]{v2}{v1}{180};
	\end{tikzpicture}
	\caption{A Feynman diagram, the fish diagram,  corresponding to $ \widetilde{\Pi}^R.$}
	\label{fig:fish}
\end{figure}
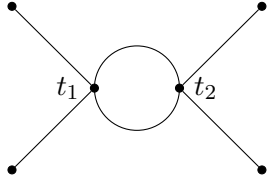

The leading order contribution to $ \widetilde{\Pi}_R(t_1,t_2) $,  and 
$ \widetilde{\kappa} (t_1)\equiv \widetilde{\kappa}(t_1,\omega=0)$,  is obtained from  the ``fish'' diagram, shown in Fig.~\ref{fig:fish}, 
 with the internal lines corresponding to two $ \eta $ or two $ \chi $ free propagators:
\begin{equation}\label{eqn:fish}
	\begin{aligned}
	     \widetilde{\Pi}_R(t_1,t_2)
	        =  &\widetilde{\Pi}_{11}(t_1,t_2)-\widetilde{\Pi}_{12}(t_1,t_2)\\
		 =  &\frac{i}{2}\sum_{\xi=\eta,\chi}\int(-ig_\xi)^2\frac{d^3q}{(2\pi)^3}\Big[(\Delta_\xi)_{11}
		     (\boldsymbol{q},t_1,t_2)(\Delta_\xi)_{11}(-\boldsymbol{q},t_1,t_2)\\
		   &  -(\Delta_\xi)_{12}(\boldsymbol{q},t_1,t_2)(\Delta_\xi)_{12}(-\boldsymbol{q},t_1,t_2)\Big]\\
		=  &\theta(t_1-t_2)\sum_{\xi=\eta,\chi}g_{\xi}^{2}\int\frac{d^3q}{(2\pi)^3}\mathrm{Im} 
		   \left[(\Delta_\xi)_>(\boldsymbol{q},t_1,t_2)^2\right].
	\end{aligned}
\end{equation}

In order to perform  the integral over 3-momentum and carry out Fourier transformation over $ t=t_1-t_2 $, we first expand the integrand in~(eq.~\ref{eqn:fish}) to  first order in $ H $,  
\begin{equation}\label{propagator squared}
	\begin{aligned}
		& (\Delta_{\xi})_>(\boldsymbol{q},t_1,t_2)^2\\
		\approx & \frac{1}{4\omega_{\xi q}^{2}}\left[(1-2Ht_1\overline{\gamma}_0)+Ht\overline{\gamma}_1+2iHt_1t\overline{\gamma}_2-iHt^2\overline{\gamma}_2\right]\left(1+f\left(\omega_{\xi q}\right)\right)^2e^{-2i\omega_{\xi q} t}\\
		&+(\omega_{\xi q}\to -\omega_{\xi q})-\frac{1}{2\omega_{\xi q}^2}\left[(1+Ht_1\gamma'_0)+Ht\gamma'_1\right](1+f(\omega_{\xi q}))f(\omega_{\xi q}),
	\end{aligned}
\end{equation}
where $ \omega_{\xi q}=\sqrt{\boldsymbol{q}^2+M_\xi^2} $ and
\begin{equation}
	\left\{
	\begin{aligned}
		& \overline{\gamma}_0(\omega_{\xi q})=3+\frac{\gamma M_\xi^2}{\omega_{\xi q}}f(\omega_{\xi q})-\frac{\omega_{\xi q}^2-M_\xi^2}{\omega_{\xi q}^2}\\
		& \overline{\gamma}_1(\omega_{\xi q})=3+2\frac{\gamma M_\xi^2}{\omega_{\xi q}}f(\omega_{\xi q})-\frac{\omega_{\xi q}^2-M_\xi^2}{\omega_{\xi q}^2}\\
		& \overline{\gamma}_2\left(\omega_{\xi q}\right)=\frac{\omega_{\xi q}^2-M_\xi^2}{\omega_{\xi q}}\\
		&\gamma'_0(\omega_{\xi q})=-6+2\frac{\omega_{\xi q}^2-M_\xi^2}{\omega^2_{\xi q}}-\left(1+2f(\omega_{\xi q})\right)\frac{\gamma M_\xi^2}{\omega_{\xi q}}\\
		& \gamma'_1(\omega_{\xi q})=3+\frac{\gamma M_\xi^2}{\omega_{\xi q}}\left(1+2f(\omega_{\xi q})\right)-\frac{\omega^2_{\xi q}-M_\xi^2}{\omega_{\xi q}^2}
	\end{aligned}
	\right. ~.
\end{equation}

For  a general function $K$ given by 
$$K=  K_0(\omega_{\xi q}) + t K_1(\omega_{\xi q}) +t^{2} K_2(\omega_{\xi q})$$ 
with  $K_0(\omega_{\xi q})$, $K_1(\omega_{\xi q}) $, $K_2(\omega_{\xi q})$ being   generic  rational functions of  
$ \omega_{\xi q} $,    
we  obtain  an  explicit   expression of  its Fourier transformation  
   in   Appendix~\ref{appendix: Fourier of general form}: 
\begin{equation}   \label{general_form_real_part}
\begin{aligned}
& \operatorname{Re} F^{\omega}\left[ \theta(t) \operatorname{Im} \int \frac{d^{3} q}{(2 \pi)^{3}}\left[K_{0}\left(\omega_{{\xi q}}\right)+t K_{1}\left(\omega_{{\xi q}}\right)+t^{2} K_{2}\left(\omega_{{\xi q}}\right)\right] e^{-2 i \omega_{{\xi q}} t}\right]\\
=&\left.I_{re}\left[K_{0}\right]\right|_{\alpha_\xi=0}+\frac{\partial}{\partial \omega} \left.I_{im}\left[K_{1}\right]\right|_{\alpha_\xi=0}-\frac{\partial^{2}}{\partial \omega^{2}} \left.I_{re}\left[K_{2}\right]\right|_{\alpha_\xi=0}~,
\end{aligned}
\end{equation}
where $ \alpha_\xi $, which will be defined and used in the following section, is related to the decay rate of a 
$\xi$ field.  $I_{r e}\left[K\left(\omega_{{\xi q}}\right)\right]$,  
and $I_{im}\left[K\left(\omega_{{\xi q}}\right)\right]$ are given as follows, 
\begin{equation}\label{definition_Ire_Iim}
	\left\{
	\begin{aligned}
		I_{r e}\left[K\left(\omega_{{\xi q}}\right)\right]=&\frac{1}{(2 \pi)^{3}} \int_{M_\xi}^{\infty} d \omega_{{\xi q}} 4 \pi \omega_{{\xi q}} \sqrt{\omega_{{\xi q}}^{2}-M_\xi^2}\Bigg\{\operatorname{Re} K\left(\omega_{{\xi q}}\right) \frac{2 \omega_{{\xi q}}}{\omega^{2}-4 \omega_{{\xi q}}^{2}}\\
		&+\operatorname{Im} K\left(\omega_{{\xi q}}\right)\left[\frac{\alpha_\xi}{2\omega_{{\xi q}}\left(\omega+2 \omega_{{\xi q}}\right)^{2}}+\frac{\alpha_\xi}{2\omega_{{\xi q}}\left(\omega-2 \omega_{{\xi q}}\right)^{2}}\right]\Bigg\}~,\\
		I_{im}\left[K\left(\omega_{{\xi q}}\right)\right]=&\frac{1}{(2 \pi)^{3}} \int_{M_\xi}^{\infty} d \omega_{{\xi q}} 4 \pi \omega_{{\xi q}} \sqrt{\omega_{{\xi q}}^{2}-M_\xi^2}\Bigg\{\operatorname{Im} K\left(\omega_{{\xi q}}\right) \frac{\omega}{\omega^{2}-4 \omega_{{\xi q}}^{2}}\\
		&-\operatorname{Re} K\left(\omega_{{\xi q}}\right)\left[\frac{\alpha_\xi}{2\omega_{{\xi q}}\left(\omega-2 \omega_{{\xi q}}\right)^{2}}-\frac{\alpha_\xi}{2\omega_{{\xi q}}\left(\omega+2 \omega_{{\xi q}}\right)^{2}}\right]\Bigg\}~.
	\end{aligned}\right.
\end{equation}

Combining~(eq.~\ref{propagator squared}) and~(eq.~\ref{definition_Ire_Iim}) we obtain,
\begin{equation}
	\begin{aligned}
		&\sum_{\xi}g_\xi^2\operatorname{Re} F^{\omega} \left[\theta(t) \operatorname{Im} \int \frac{d^{3} q}{(2 \pi)^{3}} (\Delta_{\xi})_>(\boldsymbol{q},t_1,t_2)^2\right]\\
		=&\frac{1}{2}\sum_{\xi}g_{\xi}^2\left\{I_{re}\left[\frac{1}{2\omega_{\xi q}^{2}}(1-2Ht_1\overline{\gamma}_0(\omega_{\xi q}))\left(1+f(\omega_{\xi q})\right)^2-(\omega_{\xi q}\to -\omega_{\xi q})\right]\right.\\
		&+\frac{\partial}{\partial \omega}I_{im}\left[\frac{H}{2\omega_{\xi q}^{2}}(\overline{\gamma}_1(\omega_{\xi q})+2it_1\overline{\gamma}_2(\omega_{\xi q}))\left(1+f(\omega_{\xi q})\right)^2+(\omega_{\xi q}\to -\omega_{\xi q})\right]\\
		&\left.+\frac{\partial^2}{\partial \omega^2}I_{re}\left[iH \frac{1}{2\omega_{\xi q}^{2}}\overline{\gamma}_2(\omega_{\xi q})\left(1+f(\omega_{\xi q})\right)^2-(\omega_{\xi q}\to -\omega_{\xi q})\right]\right\}~.
	\end{aligned}
\end{equation}
Setting $ \omega=0 $ and performing the integrals,  we arrive at, 
\begin{equation}\label{eqn:kappatilde}
	\begin{aligned}
		\tilde{\kappa}(t_1)&=\left.\sum_{\xi}g_\xi^2\operatorname{Re} F^{\omega} \biggg[\theta(t) \operatorname{Im} \int \frac{d^{3} q}{(2 \pi)^{3}} (\Delta_{\xi})_>(\boldsymbol{q},t_1,t_2)^2\right|_{\omega=0}\biggg]\\
		&=-\sum_{\xi=\eta,\chi}g_{\xi}^{2}(1-4Ht_1)\frac{1}{32\pi M_\xi\gamma}~.
	\end{aligned}
\end{equation}
If we set $ H=0 $, we obtain the familiar results in flat spacetime, 
\begin{equation}
	\bc(t_1)\approx -\frac{\lambda^2T}{32\pi M_\eta}-\frac{h^2T}{32\pi M_\chi}~.
\end{equation}
using Minkowski-space propagators in loop diagrams.

\subsection{Computation of \dc{1}}
\label{sec:kappa_tilde_derivative}
As mentioned in~\cite{Cheung:2015iqa}, the leading contribution to  $\dc{1}$  comes from the fish diagram, Fig.~\ref{fig:fish},  but with the one-loop corrected ($ \eta-$ or $ \chi-$) propagators.  
Such propagators rely on the decay rates, which has been calculated in~\cite{Cheung:2015iqa} in flat spacetime:
\begin{equation}\label{Gamma_in_CDKK}
\Gamma_\chi=\frac{\lambda'^2+3h^2}{256\pi^3\gamma^2\omega_\chi}\equiv \frac{\alpha_\chi}{\omega_\chi}, 
\hspace{0.8cm} 
\Gamma_\eta=\frac{\lambda^2+3h^2}{256\pi^3\gamma^2\omega_\eta}\equiv \frac{\alpha_\eta}{\omega_\eta}.
\end{equation}
In the present  situation,  since we have assumed the system to be in pseudo-equilibrium and we do not consider higher order corrections, we  replace (eq.~\ref{Gamma_in_CDKK})  by
\begin{equation}
\Gamma_\chi(t')=\frac{\lambda'^2+3h^2}{256\pi^3\beta(t')^2\Omega_\chi(t')}~,
\hspace{0.8cm} 
\Gamma_\eta(t')=\frac{\lambda^2+3h^2}{256\pi^3\beta(t')^2\Omega_\eta(t')}~.
\end{equation}

Similar to~(eq.~\ref{general_form_real_part}),  we have
\begin{equation}\label{general_form_cross_part}
\begin{aligned}
&F^{\omega}\left[\theta(t) \operatorname{Im} \int \frac{d^{3} q}{(2 \pi)^{3}}\left(K_{0}+t K_{1}+t^{2} K_{2}\right) e^{- \frac{\alpha_\xi}{\omega_{q}} t}\right]\\
=&\int \frac{d^3q}{(2\pi)^3}\left(\mathrm{Im}K_0\frac{\omega_q \alpha_\xi}{\alpha_\xi^2+\omega^2\omega^2_q}+\mathrm{Im}K_1\frac{\partial}{\partial \omega}\frac{\omega\omega^2_q}{\alpha_\xi^2+\omega^2\omega^2_q}-\mathrm{Im}K_2\frac{\partial^2}{\partial \omega^2}\frac{\omega_q \alpha_\xi}{\alpha_\xi^2+\omega^2\omega^2_q}\right)\\
&+i\int \frac{d^3q}{(2\pi)^3}\left(\mathrm{Im}K_0\frac{\omega\omega^2_q}{\alpha_\xi^2+\omega^2\omega^2_q}-\mathrm{Im}K_1\frac{\partial }{\partial \omega}\frac{\omega_q\alpha_\xi}{\alpha_\xi^2+\omega^2\omega^2_q}-\mathrm{Im}K_2\frac{\partial ^2}{\partial \omega^2}\frac{\omega\omega^2_q}{\alpha_\xi^2+\omega^2\omega^2_q}\right)
\end{aligned}
\end{equation}
and
\begin{equation}\label{general_form_imaginary_part}
\begin{aligned}
& \operatorname{Im} F^{\omega} \left\{\theta(t) \operatorname{Im} \int \frac{d^{3} q}{(2 \pi)^{3}}\left[K_{0}\left(\omega_{{\xi q}}\right)+t K_{1}\left(\omega_{{\xi q}}\right)+t^{2} K_{2}\left(\omega_{{\xi q}}\right)\right] e^{-2 i \omega_{{\xi q}} t}\right\}\\
=&I_{i m}\left[K_{0}\right]-\frac{\partial}{\partial \omega} I_{r e}\left[K_{1}\right]-\frac{\partial^{2}}{\partial \omega^{2}} I_{i m}\left[K_{2}\right],
\end{aligned}
\end{equation}
where $ I_{re} $ and $ I_{im} $ are defined in~(eq.~\ref{definition_Ire_Iim}), and the derivation of which is shown in Appendix~\ref{appendix: Fourier of general form}.  By expanding the propagator to the first order in $ H $ and making the following peak approximation, 
\begin{equation}
f(\omega_{\xi q})\to \frac{1}{\gamma \omega_{\xi q}}, \int_{m}^{\infty}\to \int_{m}^{1/\gamma}, 
\end{equation}
and then with the help of~(eq.~\ref{general_form_cross_part}) and~(eq.~\ref{general_form_imaginary_part}) to perform the Fourier transformations, to the first order in $H$  and to the lowest orders in $\alpha_\chi$ and $\alpha_\eta$ we get, 
\begin{equation}\label{kappa_tilde_first_derivative}
\begin{aligned}
&-i \dc{1}\\
=&-h^2\left[-\frac{32\pi\gamma}{\lambda'^2+3h^2}\left(1+\log\frac{M_\chi\gamma}{2}\right)+Ht_1\frac{32\pi\gamma}{3\left(\lambda'^2+3h^2\right)}\left(5+9\log\frac{M_\chi \gamma}{2}\right)-H\frac{2^{13}\pi^4\gamma^2}{\left(\lambda'^2+3h^2\right)^2}\right]\\
&-\lambda^2\left[-\frac{32\pi\gamma}{\lambda^2+3h^2}\left(1+\log\frac{M_\chi\gamma}{2}\right)+Ht_1\frac{32\pi\gamma}{3\left(\lambda^2+3h^2\right)}\left(5+9\log\frac{M_\chi\gamma}{2}\right)-H\frac{2^{13}\pi^4\gamma^2}{\left(\lambda^2+3h^2\right)^2}\right]~.
\end{aligned}
\end{equation}

\subsection{Computation of $\ds{1}$}
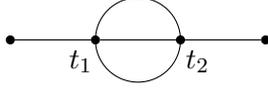
\begin{figure}
	\centering
	\begin{tikzpicture}
	\coordinate[vertex](v1);
	\coordinate[vertex,right=of v1](v2);
	\coordinate[vertex,left=of v1](e1);
	\coordinate[vertex,right=of v2](e2);
	\draw[boson](e1)--(v1) node [align=left,below]{$ t_1\quad$};
	\draw[boson](v2)node[align=right,below]{$\quad\, t_2 $}--(e2);
	\draw[boson](v1)--(v2);
	\semiloop[boson]{v1}{v2}{0};
	\semiloop[boson]{v2}{v1}{180};
	\end{tikzpicture}
	\caption{A Feynman  diagram, the sunset diagram, corresponding to the imaginary part of $ \Pi^R $}
	\label{fig:sunset}
\end{figure}
$ \ds{1} $ is determined by the imaginary part of the self-energy~\cite{Cheung:2015iqa} whose leading order contribution, for soft momenta, comes from the sunset diagram in Fig.~\ref{fig:sunset}.
\begin{equation}
	\begin{aligned}
		&\Pi^R(t_1,t_2)\\
		=&h^2\theta(t_1-t_2)\int \frac{d^3 k}{(2\pi)^3}\frac{d^3l}{(2\pi^3)}\mathrm{Im}\left[(\Delta_\chi)_>(\boldsymbol{k},t_1,t_2)(\Delta_\chi)_>(\boldsymbol{l},t_1,t_2)(\Delta_\eta)_>(\boldsymbol{k}+\boldsymbol{l},t_1,t_2)\right]~.
	\end{aligned}
\end{equation}

In Appendix~\ref{appendix: C3} we show  that we can express the derivative of $ \kappa(-\omega) $ in the following form, where the derivatives of $ I_s, I_c, J_s, J_c $ are given by equations~(\ref{I_s}), (\ref{I_c}), (\ref{J_s}),  (\ref{J_c}),
\begin{equation}\label{self_imaginary}
	\begin{aligned}
		& -i\ds{1}\\
		=&-\frac{h^2}{8}\Bigg\{ \left( 1 - 9 H t _ { 1 } \right) \left. \frac { \partial } { \partial \omega } I _ { s }  \left[ \gamma _ { 0 } \right] \right| _ { \omega = 0 } + H t _ { 1 }\left.  \frac { \partial } { \partial \omega } I _ { s } \left[ \gamma _ { 1 } \right] \right| _ { \omega = 0 } - H t _ { 1 }\left.  \frac { \partial } { \partial \omega } I _ { s } \left[ \gamma _ { 2 } \right] \right| _ { \omega = 0 } - 2 H t _ { 1 }\left.  \frac { \partial ^ { 2 } } { \partial \omega ^ { 2 } } I _ { c } \left[ \gamma _ { 3 } \right] \right|_ { \omega = 0 }\\
		& -\frac { 9 } { 2 } H\left.  \frac { \partial } { \partial \omega } J _ { s } \left[ \gamma _ { 0 } \right] \right| _ { \omega = 0 } - \frac { 1 } { 2 } H \left. \frac { \partial } { \partial \omega } J _ { s } \left[ \gamma _ { 1 } \right] \right|_ { \omega = 0 } + H \left. \frac { \partial } { \partial \omega } J _ { s } \left[ \gamma _ { 2 } \right] \right| _ { \omega = 0 } - H\left.  \frac { \partial } { \partial \omega } J _ { c } \left[ \gamma _ { 3 } \right] \right|_ { \omega = 0 }\Bigg\}~.
	\end{aligned}
\end{equation}
This  derivation  is not much different from the calculation of the imaginary part of the self-energy performed  in~\cite{Cheung:2015iqa}.   But  the analogous integrals in de Sitter spacetime  are more complicated:  we do not have delta functions resulted from  momentum conservation,  which in turn greatly  simplify  the  subsequent calculations.     
Our strategy is to  calculate each term in the above equation,  with the assumption that $  M_\chi/M_\eta\ll 1 $,  to obtain analytical results of the integrals. 

Let us now  calculate  the first line in the curly bracket in~(eq.~\ref{self_imaginary}),  which is dominated by regions where $ \omega_{\chi k},\omega_{\chi l}\ll 1/\gamma $, since in these regions the Bose distribution function has a peak.   We  can then  make an  approximation, $ f(x)\approx 1/(\gamma x)-1/2 \quad (\text{for } x\ll 1/\gamma)$, to simplify the integrals. Since $ \gamma_0,\gamma_1,\gamma_2 $ ( at least to the zeroth order in $ M_\eta\gamma $) do not change when all three arguments change simultaneously,  while $ \gamma_3 $ changes sign, the temperature dependent part in $ \partial /\partial \omega I_s[\gamma_{0,1,2}]|_{\omega=0} $ and in $ \partial^2 /\partial^2 \omega I_c[\gamma_3] |_{\omega=0}$ appears  in the following forms:
\begin{equation}
	\begin{aligned}
		& \left(1+f(\omega_{\eta q})\right)f(\omega_{\chi k})f(\omega_{\chi l})-f(\omega_{\eta q})\left(1+f(\omega_{\chi k})\right)\left(1+f(\omega_{\chi l})\right)\approx \frac{1}{\gamma^2}\left(\frac{1}{\omega_{\chi k}\omega_{\chi l}}-\frac{1}{\omega_{\eta q}\omega_{\chi k}}-\frac{1}{\omega_{\eta q}\omega_{\chi l}}\right),\\
		& \left(1+f(\omega_{\eta q})\right)f(\omega_{\chi k})f(\omega_{\chi l})+f(\omega_{\eta q})\left(1+f(\omega_{\chi k})\right)\left(1+f(\omega_{\chi l})\right)\approx \frac{1}{\gamma^3\omega_{\eta q}\omega_{\chi k}\omega_{\chi  l}}~. 
	\end{aligned}
\end{equation}
We thus  conclude that $ \partial /\partial \omega I_s[\gamma_{0,1,2}]|_{\omega=0} $ is much smaller than $ \partial^2 /\partial^2 \omega I_c[\gamma_3] |_{\omega=0}$ and  the former can be safely neglected. 

Considering $- \left. \frac { \partial ^ { 2 } } { \partial \omega ^ { 2 } } I _ { c } \left[ \gamma _ { i } \right] \right|_ { \omega = 0 }$  it  is the sum of the following three integrals~(eq.~\ref{I_c}),
\begin{equation}\label{ic1}
	\begin{aligned}
		&- \frac { 1 } { 2 ( 2 \pi ) ^ { 3 } } \int _ { M_\chi } ^ { \infty } d \omega_{\chi l}\int _ { M_\chi t _ { 0 m } } ^ { M_\chi t _ { 0 p } } d \omega_{\chi k} \frac { \partial ^ { 2 } } { \partial \omega ^ { 2 } }\Big[\gamma _ { i } \left( - \omega + \omega_{\chi k} + \omega_{\chi l}- \omega_{\chi k} , - \omega_{\chi l} \right) \left( 1 + f \left( \omega_{\chi k} + \omega_{\chi l} - \omega \right) \right) f \left( \omega_{\chi k} \right) f \left( \omega_{\chi l} \right)\\
		& - \gamma _ { i } \left( \omega - \omega_{\chi k} - \omega_{\chi l} , \omega_{\chi k} , \omega_{\chi l} \right) f \left( \omega_{\chi k} + \omega_{\chi l} - \omega \right) \left( 1 + f \left( \omega_{\chi k} \right) \right) \left( 1 + f \left( \omega_{\chi l} \right) \right)\Big]\bigg|_{\omega=0},
	\end{aligned}
\end{equation}
\begin{equation}\label{ic2}
	\begin{aligned}
		& - \frac { 1 } { 2 ( 2 \pi ) ^ { 3 } } \int _ { M_\chi } ^ { \infty } d \omega_{\chi l}M_\chi^2(t_{1p})^2\frac{\partial}{\partial \omega_{\chi k}}\Big[\gamma_i(\omega_{\chi k}+\omega_{\chi l},-\omega_{\chi k},-\omega_{\chi l})\left(1+f\left(\omega_{\chi k}+\omega_{\chi l}\right)\right)f\left(\omega_{\chi k}\right)f\left(\omega_{\chi l}\right)\\
		&-\gamma_i\left(-\omega_{\chi k}-\omega_{\chi l},\omega_{\chi k},\omega_{\chi l}\right)f\left(\omega_{\chi k}+\omega_{\chi l}\right)\left(1+f\left(\omega_{\chi k}\right)\right)\left(1+f\left(\omega_{\chi l}\right)\right)\Big]\bigg|_{\omega_{\chi k}\to M_\chi t_{0p}},
	\end{aligned}
\end{equation}
\begin{equation}\label{ic3}
	\begin{aligned}
		&\frac { 1 } { 2 ( 2 \pi ) ^ { 3 } } \int _ { M_\chi } ^ { \infty } d \omega_{\chi l}M_\chi^2(t_{1m})^2\frac{\partial}{\partial \omega_{\chi k}}\Big[\gamma_i(\omega_{\chi k}+\omega_{\chi l},-\omega_{\chi k},-\omega_{\chi l})\left(1+f\left(\omega_{\chi k}+\omega_{\chi l}\right)\right)f\left(\omega_{\chi k}\right)f\left(\omega_{\chi l}\right)\\
		&-\gamma_i\left(-\omega_{\chi k}-\omega_{\chi l},\omega_{\chi k},\omega_{\chi l}\right)f\left(\omega_{\chi k}+\omega_{\chi l}\right)\left(1+f\left(\omega_{\chi k}\right)\right)\left(1+f\left(\omega_{\chi l}\right)\right)\Big]\bigg|_{\omega_{\chi k}\to M_\chi t_{0m}}~.
	\end{aligned}
\end{equation}
It is easy to check that $ M_\chi t_{0p} $ grows rapidly with $ \omega_{\chi l} $:  $ M_\chi t_{0p}\sim\frac{M_\eta^2}{M_\chi^2}\omega_{\chi l} $ while $ M_\chi t_{0m} $ decreases with $ \omega_{\chi l} $. Therefore for the integral~(eq.~\ref{ic2}) the rational-function approximation of the Bose distribution function is inappropriate.  The integrand will indeed be negligible due to the large exponential in the denominator.  
On the other hand, it is viable to use the rational-function approximation in the integral~(eq.~\ref{ic3}),  which turns out to be much larger than~(eq.~\ref{ic2}). For the integral~(eq.~\ref{ic1}), we can see that it would not be of much difference in the order of magnitude if we replace the partial derivative $ \partial /\partial \omega $ by $ \partial /\partial \omega_{\chi k} $. 
After carrying out this replacement, it can be written as $ [(\ref{ic2})/(t_{1p})^2-(\ref{ic3})/(t_{1m})^2]$.  The first term can be neglected because~(eq.~\ref{ic2}) is small and $ t_{1p} $ is large. We can  also neglect the second term because $ t_{1m} $ is much greater than 1 when $ \omega_{\chi l} $ is small, and we can make a rough estimate of the ratio of the contribution of $ t_{1m} $ to that of $ 1 $,
\begin{equation}
	\dfrac{\displaystyle\int_{m}^{M}t_{1m}d\omega_{\chi l}}{\displaystyle\int_{M_\chi}^{M_\eta}d\omega_{\chi l}}\approx \frac{M_\eta}{6M_\chi}~.
\end{equation}
It shows that~(eq.~\ref{ic1}) is indeed much smaller than~(eq.~\ref{ic3}) and therefore we will only keep~(eq.~\ref{ic3}).

With the  above approximations performed~(eq.~\ref{I_c}) can be integrated out:
\begin{equation}\label{I_result}
	\begin{aligned}
		&- \left. \frac { \partial ^ { 2 } } { \partial \omega ^ { 2 } } I _ { c } \left[ \gamma _ { 3} \right] \right|_ { \omega = 0 }\\
		\approx&\frac { 1 } {  ( 2 \pi ) ^ { 3 } }\int _ { M_\chi } ^ { \infty } d \omega_{\chi l}M_\chi ^ { 2 } \left( t _ { 1 m } \right) ^ { 2 } \frac { \partial } { \partial \omega_{\chi k} } \left[ \gamma _ { 3 } \left( \omega_{\chi k} + \omega_{\chi l} , - \omega_{\chi k} , - \omega_{\chi l} \right)\frac{1}{\gamma^3\left(\omega_{\chi k}+\omega_{\chi l}\right)\omega_{\chi k}\omega_{\chi l}}\right]\bigg|_{\omega_{\chi k}\to M_\chi t_{0m}}\\
		\approx &\frac{T^3}{4\pi^3M_\eta^2}\left[\frac{M_\eta^8}{(M_\eta^4+4M_\chi^2T^2)^2}-\frac{M_\eta^4}{M_\eta^4+4M_\chi^2T^2}-\frac{4M_\chi^2M_\eta^4T^2}{(M_\eta^4+4M_\chi^2T^2)^2}+\frac{8M_\chi^4T^2}{M_\eta^2(M_\eta^4+4M_\chi^2T^2)}\right.\\
		&\left.+\log\left(\frac{1}{4}\right)-1+\log\left(\frac{M_\eta^6}{M_\chi^4\sqrt{M_\eta^4+4M_\chi^2T^2}}\right)\right]~.
	\end{aligned}
\end{equation}
The accuracy of this approximation can be seen in  some examples in Table~\ref{table:icd},   the error is around 3\%.
\begin{table}
	\begin{center}
	\begin{tabular}{|c|c|c|c|c|c|}
			\hline
			$ T $ & $ M_\eta $ & $ M_\chi $ & analytic result & numerical result & error\\ \hline
			$ 1\times 10^5 $ & $ 500 $ & $ 1 $ & $7.01\times10^{8}$ & $ 7.30\times10^8 $ & $-4\%$\\ \hline
			$ 1\times 10^5 $ & $ 1000 $ & $ 1 $ & $ 2.03\times 10^8 $ & $ 1.99\times 10^8 $ & 2\%\\ \hline
			$ 1\times 10^4 $ & $ 100 $ & $ 1 $ & $ 1.20\times 10^7 $ & $ 1.19\times 10^7 $ & $ 0.9\% $\\ \hline
			$ 1\times 10^4 $ & $ 100 $ & $ 0.5 $ & $ 1.45\times 10^7 $ & $ 1.51\times 10^7 $ & $ -3.8\% $\\ \hline
	\end{tabular}
	\caption{Comparison of analytic and numerical results: here we choose some numerical values for the parameters and list the analytic results, which is determined by~(eq.~\ref{I_result}), and the numerical results, which is numerically computed from the first line in the curly bracket in~(eq.~\ref{self_imaginary}), corresponding to each of the choices.}
	\label{table:icd}
	\end{center}
\end{table}	

\begin{figure}[h]
	\begin{center}
		\includegraphics[width=15cm]{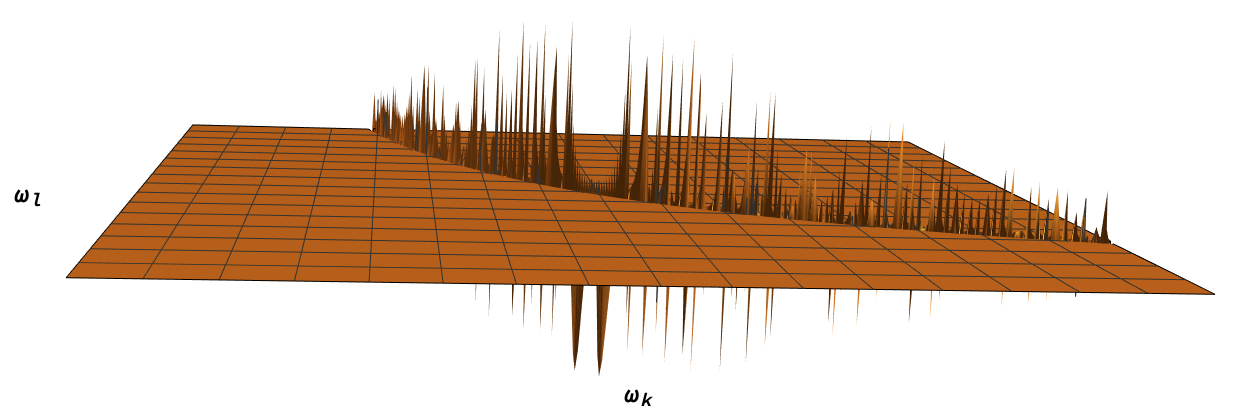}
	\end{center}
	\caption{Plot of the integrand in~(eq.~\ref{principle_gamma3}). }   \label{principle_plot}
\end{figure}

Next we compute $ \partial /\partial \omega J_c[\gamma_3]|_{\omega=0} $ and $ \partial /\partial \omega J_s[\gamma_{0,1,2}]|_{\omega=0} $, for which we assume that: $\lambda, \lambda', h\ll M_\chi \gamma$.  
As we have mentioned in Section~\ref{sec:assumptions}, in order for the calculations to be carried out perturbatively, $ \lambda,\lambda', h $ should be small enough and, in particular,  smaller than any dimensionless quantity that can be constructed using dimensionful dynamical quantities in the model.   There are eight terms in $\partial /\partial \omega J_c[\gamma_i]|_{\omega=0}$, with different signs $ \chi_q,\chi_k,\chi_l $ before $ \omega_{\eta q},\omega_{\chi k},\omega_{\chi l} $. We first compute the terms in $\partial /\partial \omega J_c[\gamma_3]|_{\omega=0}$ that correspond to $ \chi_q=1,\chi_k=-1,\chi_l=-1 $ and $ \chi_q=-1,\chi_k=1,\chi_l=1 $:
\begin{equation}\label{principle_gamma3}
   \begin{aligned}
	& 6 \frac { 1 } { ( 2 \pi ) ^ { 4 } } \int _ { M_\chi } ^ { \infty } d \omega_{\chi k} d \omega_{\chi l} \int _ { - 1 } ^ { 1 } d w \frac { \sqrt { \omega_{\chi k} ^ { 2 } - M_\chi ^ { 2 } } \sqrt { \omega_{\chi l} ^ { 2 } - M_\chi ^ { 2 } }  \left( \omega _ { \eta q } -\omega_{\chi k} - \omega_{\chi l} \right) ^ 4} { \omega _ { \eta q } \left[\left( \omega _ { \eta q } - \omega_{\chi k} - \omega_{\chi l} \right) ^2+\alpha_\chi^2\left(1/(2\omega_{\chi k})+1/(2\omega_{\chi l})\right)^2\right]  ^4} \gamma _ { 3} \left( \omega _ { \eta q } , -\omega_{\chi k} , -\omega_{\chi l} \right)\\
	&\times [(1+f(\omega_{\eta q}))(1+f(-\omega_{\chi k}))(1+f(-\omega_{\chi l}))+(1+f(-\omega_{\eta q}))(1+f(\omega_{\chi k}))(1+f(\omega_{\chi l}))]~.
   \end{aligned}
\end{equation}
These  are the dominant contribution to the principle part of the sunset diagram.
The second line in the above formula has a peak when $\omega_{\chi k},\omega_{\chi l}\ll 1/\gamma$,  allowing us  to approximate it by, 
\begin{equation}
	\frac{1}{4}+\frac{1}{\gamma^2}\left(\frac{1}{\omega_{\chi k} \omega_{\chi l}}
	-\frac{\omega_{\chi k}+\omega_{\chi l}}{\omega_{\chi k} \omega_{\chi l} \omega_{\eta q}}\right)~,
\end{equation}
and to cut down the upper limit of integration to $ 1/\gamma $.
Let us now consider the following term in the integral:
\begin{equation}\label{fraction_part}
	G(\omega_{\chi k},\omega_{\chi l},w)=\frac{g(\omega_{\chi k},\omega_{\chi l},w)^4}{\left[g(\omega_{\chi k},\omega_{\chi l},w)^2+\alpha_\chi^2\left(\dfrac{1}{2\omega_{\chi k}}+\dfrac{1}{2\omega_{\chi l}}\right)^2\right]^4},
\end{equation}
where
\begin{equation}
	\begin{aligned}
		g(\omega_{\chi k},\omega_{\chi l},w)&\equiv \omega_{\eta q}-\omega_{\chi k}-\omega_{\chi l}\\
		&=\sqrt{\omega_{\chi k}^2+\omega_{\chi l}^2+2w\sqrt{\omega_{\chi k}^2-M_\chi^2}\sqrt{\omega_{\chi l}^2-M_\chi^2}+M_\eta^{2}-2M_\chi^2}-\omega_{\chi k}-\omega_{\chi l}.
	\end{aligned}
\end{equation}
Since $ \alpha_\chi/M_\chi^2\ll 1/(256\pi^3)\Rightarrow \alpha_\chi/M_\chi^2\ll 10^{-4} $,  one infers that $ G $ has a peak (see Fig.~\ref{principle_plot} for the plot of the integrand in~(eq.~\ref{principle_gamma3})) around some points at which $ g(\omega_{\chi k},\omega_{\chi l},w) $ and $ \alpha_\chi(1/\omega_{\chi k}+1/\omega_{\chi l}) $ have the same order of magnitude.  Hence, 
\begin{equation}\label{order_nu}
	\begin{aligned}
    & \omega_{\chi k}\omega_{\chi l}=\frac{M_\eta^2}{2(1-w)}+O\left(M_\chi^2\right)+O\left(\alpha_\chi\right),\\
		& \omega_{\chi k},\omega_{\chi l}=O(M_\eta)+O\left(\frac{M_\chi^2}{M_\eta}\right)   +O\left(\frac{\alpha_\chi}{M_\eta}\right).
	\end{aligned}
\end{equation}
Furthermore  we note that if we vary $ g(\omega_{\chi k},\omega_{\chi l}, w) $ by an amount of order $ \alpha_\chi/M_\eta $, then $ \alpha_\chi(1/\omega_{\chi k}+1/\omega_{\chi l}) $ would vary by an amount of order                          $\alpha_\chi^2/M_\eta^3 $,  i.e.,  the latter can be regarded as constant.    Hence, to determine the local maximum points of $ G(\omega_{\chi k},\omega_{\chi l},w) $ to an  order of $ O(\alpha_\chi/M_\eta) $, we  require that 
\begin{equation}\label{eqn:max_point}
	g(\omega_{\chi k},\omega_{\chi l}, w)=\alpha_\chi\left(\frac{1}{2\omega_{\chi k}}+\frac{1}{2\omega_{\chi l}}\right)~.
\end{equation}
$ \omega_{\chi k}, \omega_{\chi l} $ are apparently symmetric in these formulae.  
To simplify the calculations we define a set of new variables: 
$ \mu=\omega_{\chi k}+\omega_{\chi l}, \nu=\omega_{\chi k}\omega_{\chi l}/M_\eta $. 
Then~(eq.~\ref{eqn:max_point}) becomes
\begin{equation}
	g(\mu,\nu)=\frac{\alpha_\chi\mu}{2M_\eta\nu},
\end{equation}
and to lowest order in $ \alpha_\chi/M_\eta $ and $ M_\chi^2/M_\eta $, the solution to the above equation is
\begin{equation}
	\nu_{max}=\frac{M_\eta}{2(1-w)}~.
\end{equation}

We will expand $ G $ near $ \nu_{max} $ at which $ G $ has a peak and then the integration region becomes a band that is narrow in the $ \nu $ direction and runs along the $ \mu $ direction. From~(eq.~\ref{order_nu}) we see that the width of the band can be chosen to be $ \alpha_\chi/M_\chi $ since $ \alpha_\chi/M_\chi\gg \alpha_\chi/M_\eta $.
First we expand the function $ g $ around $ \nu_{max} $:
\begin{equation}
	g\left(\mu,\nu_{max}(\mu)+\Delta\right)=g\left(\mu,\nu_{max}(\mu)\right)+\Delta\cdot \left.\frac{\partial g}{\partial \nu}\right|_{\mu,\nu_{max}(\mu)}=\alpha_\chi\frac { \mu } { 2M_\eta \nu_ { \max } ( \mu ) } - \frac { M_\eta } { \mu } ( 1 - w ) \Delta.
\end{equation}
Inserting it into $ G $ we have
\begin{equation}
	G \left( \mu , \nu_ { max } ( \mu ) + \Delta \right)=\frac { g \left( \mu , \nu_ { max } ( \mu ) + \Delta \right) ^ { 4 } } { \left[ g \left( \mu , \nu_ { max } ( \mu ) + \Delta \right) ^ { 2 } + \dfrac { \alpha_\chi ^ { 2 } \mu ^ { 2 } } { 4M _\eta^ { 2 } \nu_ { max } ( \mu ) ^ { 2 } } \right] ^ { 4 } },
\end{equation}
where we have made an approximation: $ \alpha_\chi^2 \mu^2/\left\{M_\eta^2\left[\nu_{max}(\mu)+\Delta\right]^2\right\}\approx \alpha_\chi^2 \mu^2/\left[M_\eta^2\nu_{max}(\mu)^2\right]$. The integrand in~(eq.~\ref{principle_gamma3}) becomes
\begin{equation}\label{eq: intermediate}
	6 \frac { 1 } { ( 2 \pi ) ^ { 4 } }\frac { \sqrt { \omega_{\chi k} ^ { 2 } - M_\chi ^ { 2 } } \sqrt { \omega_{\chi l} ^ { 2 } - M_\chi ^ { 2 } } } { \omega _ { \eta q } } G \left( \mu , \nu _ { max } ( \mu ) + \Delta \right)\left( \frac { \omega _ { \eta q } ^ { 2 } - M_\eta ^ { 2 } } { 2 \omega _ { \eta q } } - \frac { 1 } { 2 } \mu + \frac { M_\chi ^ { 2 } \mu } { 2 M_\eta \nu} \right) \left[ \frac { 1 } { 4 } + \frac { 1 } { \gamma ^ { 2 } M_\eta \nu } \left( 1 - \frac { \mu } { \omega _ { \eta q } } \right) \right].
\end{equation}
Because in the narrow band we have:
	$$\nu =\frac { M_\eta } { 2 ( 1 - w ) } + O \left( \frac { M_\chi ^ { 2 } } { M_\eta } \right) + O \left( \frac { \alpha_\chi } { M_\eta } \right) , $$
	$$\omega _ { \eta q } = \mu + g \left( \mu ,\nu _ { max } ( \mu ) + \Delta \right), \quad $$
	$$\frac { 1 } { \gamma ^ { 2 } M_\eta\nu} \left( 1 - \frac { \mu } { \omega _ { \eta q } } \right) \ll \frac { 1 } { 4 }.$$
Eq.~(\ref{eq: intermediate}) can be approximated by
\begin{equation}
	6 \cdot \frac { 1 } { ( 2 \pi ) ^ { 4 } } \frac { M_\eta ^ { 2 } } { 8 ( 1 - w ) } \frac { 1 } { \mu } \left[ - \frac { M_\eta ^ { 2 } } { 2 \mu } + ( 1 - w ) \frac { M_\chi ^ { 2 } \mu } { M_\eta ^ { 2 } } \right]\frac { g \left( \mu , \nu _ { max } ( \mu ) + \Delta \right) ^ { 4 } } { \left[ g \left( \mu , \nu _ { max } ( \mu ) + \Delta \right) ^ { 2 } + \dfrac { \alpha_\chi ^ { 2 } \mu ^ { 2 } } { 4M_\eta ^ { 2 } \nu _ { max } ( \mu ) ^ { 2 } } \right] ^ { 4 } }.
\end{equation}
Since outside of the band the integrand is negligible, the integration region of $ \Delta $ can be chosen to be $ (-\infty,\infty) $. We also have restrictions on the range of $ \mu $ and $ w $: in the band we have
\begin{equation}
	\mu ^ { 2 } = \left( \omega_{\chi k} + \omega_{\chi l} \right) ^ { 2 } \geq 4 \omega_{\chi k} \omega_{\chi l} \approx \frac { 2 M_\eta ^ { 2 } } { 1 - w } \geq M_\eta ^ { 2 }
	 ~~~\Rightarrow~~~ (\mu \geq M_\eta )~~\&~~(w \leq 1 - \frac { 2 M_\eta ^ { 2 } } { \mu ^ { 2 } }).
\end{equation}
Therefore the measure becomes
\begin{equation}
	\int _ { - 1 } ^ { 1 } d w \int _ { M_\chi } ^ { 1 / \gamma } d \omega_{\chi k} \int _ { M_\chi } ^ { 1 / \gamma } d \omega_{\chi l} = \int _ { - 1 } ^ { 1 - \frac { 2 M_\eta ^ { 2 } } { \mu ^ { 2 } } } d w \int _ { M_\eta } ^ { 2 / \gamma } d \mu \int _ { -\infty} ^ { \infty } d \Delta \frac { M_\eta } { \sqrt { \mu ^ { 2 } - 4 M_\eta v } }.
\end{equation}

Hence the integral (eq.~\ref{principle_gamma3}) becomes
\begin{eqnarray}\label{J_result}
	\begin{aligned}
		& 6 \cdot \frac { 1 } { ( 2 \pi ) ^ { 4 } } \int _ { - 1 } ^ { 1 - \frac { 2 M_\eta ^ { 2 } } { \mu ^ { 2 } } } d w \int _ { M_\eta } ^ { 2 / \gamma } d \mu \frac { M_\eta } { \sqrt { \mu ^ { 2 } - \dfrac { 2 M_\eta ^ { 2 } } { 1 - w } } } \frac { M_\eta ^ { 2 } } { 8 ( 1 - w ) } \frac { 1 } { \mu } \left[ - \frac { M_\eta ^ { 2 } } { 2 \mu } + ( 1 - w ) \frac { M_\chi^ { 2 } \mu } { M_\eta ^ { 2 } } \right]\\
		&\times \left[ \frac { 2M_\eta \nu _ { max } ( \mu ) } { \alpha_\chi \mu } \right] ^ { 4 } \int _ { - \infty} ^ { \infty } d \Delta\frac { \left[ 1 - \dfrac { 2( 1 - w ) M_\eta ^ { 2 } \nu _ { max } ( \mu ) } { \alpha_\chi \mu ^ { 2 } } \Delta \right] ^ { 4 } } { \left\{ \left[ 1 - \dfrac { 2( 1 - w ) M _\eta^ { 2 } \nu _ { max } ( \mu ) } { \alpha_\chi \mu ^ { 2 } } \Delta \right] ^ { 2 } + 1 \right\} ^ { 4 } }\\
		=&6 \cdot \frac { 1 } { ( 2 \pi ) ^ { 4 } } \int _ { - 1 } ^ { 1 - \frac { 2 M_\eta ^ { 2 } } { \mu ^ { 2 } } } d w \int _ { M_\eta} ^ { 2 / \gamma } d \mu \frac { M_\eta } { \sqrt { \mu ^ { 2 } - \dfrac { 2 M_\eta ^ { 2 } } { 1 - w } } } \frac { M_\eta ^ { 2 } } { 8 ( 1 - w ) } \frac { 1 } { \mu } \left[ - \frac { M _\eta^ { 2 } } { 2 \mu } + ( 1 - w ) \frac { M_\chi ^ { 2 } \mu } { M_\eta ^ { 2 } } \right]\\ 
	&{\displaystyle\times \left[ \frac { 2M_\eta \nu _ { max } ( \mu ) } { \alpha_\chi \mu } \right] ^ { 4 } \frac { \pi } { 16 } \frac { \alpha_\chi \mu ^ { 2 } } {2 ( 1 - w ) M_\eta ^ { 2 } \nu _ { max } ( \mu ) }}\\  \\
		\approx& {\displaystyle {-\frac{3\cdot 2^{21} \pi^6 \gamma^2 M_\eta^2}{35(\lambda'^2+3h^2)^3}}}~, \\
	\end{aligned}
\end{eqnarray}
where we have kept only terms  of the lowest order in $\alpha_\chi$.

Now  let us turn our attention to  the other six terms in the expression for 
$ \partial /\partial \omega J_c[\gamma_3]|_{\omega=0} $.  It is easy to check that for any of these terms with signs    $\chi_q,\chi_k,\chi_l $, the equation,
\begin{equation}
	\chi_q\omega_{\eta q}+\chi_k\omega_{\chi k}+\chi_l\omega_{\chi l}=\alpha_\chi\left(\frac{1}{2\omega_{\chi k}}+\frac{1}{2\omega_{\chi l}}\right)~,
\end{equation}
has no solution.  The integrands in these terms do not have a sharp peak and they are much smaller than the two terms we have calculated,  and can henceforth  be neglected. 
In addition,  $ \partial /\partial \omega J_s[\gamma_i]|_{\omega=0} (i=0,1,2)$ can be neglected as well.   The corresponding $ G(\omega_{\eta q},\omega_{\chi k},w) $ in these terms is of order $M_\eta^3/\alpha_\chi^3$ at the peak and is much smaller than~(eq.~\ref{fraction_part}) which is of order $ M_\eta^4/\alpha_\chi^4 $.

Combining equations~(\ref{I_result}) and~(\ref{J_result}) we obtain the derivative of the self-energy:
\begin{equation}\label{sunset_result}
	\begin{aligned}
		&-i\ds{1}\\
		=&-\frac { h ^ { 2 } } { 2 ( 4 \pi ) ^ { 3 } } \frac { 1 } { M_\eta \gamma ^ { 2 } } \left( 1 + \log \frac { M_\eta } { M_\chi } \right)-H\frac{3\cdot 2^{18} \pi^6h^2 \gamma^2 M_\eta^2}{35(\lambda'^2+3h^2)^3}\\
		&-h^2Ht_1\frac{1}{16\pi^3M_\eta^2\gamma^3}\biggg[\frac{M_\eta^8}{(M_\eta^4+4M_\chi^2/\gamma^2)^2}-\frac{M_\eta^4}{M_\eta^4+4M_\chi^2/\gamma^2}-\frac{4M_\chi^2M_\eta^4/\gamma^2}{(M_\eta^4+4M_\chi^2/\gamma^2)^2}\\
		&+\frac{8M_\chi^4/\gamma^2}{M_\eta^2(M_\eta^4+4M_\chi^2/\gamma^2)}+\log\left(\frac{1}{4}\right)-1+\log\left(\frac{M_\eta^6}{M_\chi^4\sqrt{M_\eta^4+4M_\chi^2/\gamma^2}}\right)\biggg].
	\end{aligned}
\end{equation}
Even if the Hubble parameter is very small, the second term at the right-hand side might be larger than the first term     if the coupling constants $ \lambda' $ and $ h $ are small enough. This is due to the resonance effect which amplifies the curved-spacetime effects.  The implication of this resonance on matter creation in the early universe shall be explored in a forthcoming article. 

\subsection{Computation of $\ds{2}$ and $\dc{2}$}
The contribution to $\ds{2}$ can be determined by the tadpole diagram:  
\begin{equation}\begin{aligned}
		\ds{2}&=\left[\frac{\partial^2}{\partial\omega^2}\int_{- \infty}^{\infty}dt\Pi^R(t_1,t_1-t)e^{i\omega t}\right]_{\omega=0}\\
		&=-\int_{-\infty}^{\infty}dt\Pi^R(t_1,t_1-t)t^2\\
		&\propto\int_{-\infty}^{\infty}dt\delta(t)t^2=0~.
	\end{aligned}
\end{equation}
As  for $ \dc{2} $,  the calculation is similar to that of  $ \tilde{\kappa}(t_1,\omega=0) $  and the result is
\begin{equation}\label{kappa_tilde_second_derivative}
	\dc{2}= -\left(1-4 H t_{1}\right) \frac{1}{256 \pi \gamma} \sum_{\xi=\eta,\chi} \frac{g_{\xi}^{2}}{M_{\xi}^{3}}~.
\end{equation}

\paragraph{Summary:}
Let us summarise our main results here.  The  equation of motion for $\varphi$ in an abstract form is given above in~(\ref{eqn: effective_EoM}):
\begin{equation} 
	\begin{aligned}
		&\ddot{\varphi}(t_1)+\Bigg\{\left[m_\phi^2+(1+3Ht_1)\bs(t_1,\omega=0)-3H\left(-i\ds{1}\right)\right]\varphi(t_1)\\
		&\hphantom{\ddot{\varphi}(t_1)+}
		+\frac{1}{6}\left[\lambda+(1+3Ht_1)\bc(t_1,\omega=0)-3H\left(-i\dc{1}\right)\right]\varphi(t_1)^3\Bigg\}\\
		&+\Bigg\{3H+(1+3Ht_1)\left(-i\ds{1}\right)+3H\ds{2}\\
		&\hphantom{+}
		+\frac{1}{3}\left[(1+3Ht_1)\left(-i\dc{1}\right)+3H\dc{2}\right]\varphi(t_1)^2\Bigg\}\,\dot{\varphi}(t_1) =0~.
	\end{aligned}
\end{equation}
Using the  results obtained in this section we obtain an explicit expression of the equation of motion  for $ \varphi $, describing its dynamics in an expanding or a contracting cosmic background:
\begin{equation}\label{eqn: final_EoM}
\begin{aligned}
&\ddot{\varphi}(t_1)+\biggg\{\left[m_\phi^2+\left(1+Ht_1\right)\frac { \lambda + h } { 24 \gamma ^ { 2 } }+H\frac {3} { 2 ( 4 \pi ) ^ { 3 } } \frac { h ^ { 2 } } { M_\eta \gamma ^ { 2 } } \left( 1 + \log \frac { M_\eta } { M_\chi } \right)\right]\varphi(t_1)\\
&\hphantom{\ddot{\varphi}(t_1)+}
+\Bigg[\lambda-\left(1-Ht_1\right)\left(\frac{\lambda^2}{32\pi M_\eta \gamma}+\frac{h^2}{32\pi M_\chi \gamma}\right)\\
&\hphantom{\ddot{\varphi}(t_1)+}-3H\bigg(\frac{32\pi\gamma h^2}{\lambda'^2+3h^2}+\frac{32\pi\gamma \lambda^2}{\lambda^2+3h^2}\bigg)\left(1+\log\frac{M_\chi\gamma}{2}\right)\Bigg]\frac{\varphi(t_1)^3}{6}\biggg\}\\
&+\biggg\{-\frac { h ^ { 2 } } { 2 ( 4 \pi ) ^ { 3 } } \frac { 1 } { M_\eta \gamma ^ { 2 } } \left( 1 + \log \frac { M_\eta } { M_\chi } \right)+3H-H\frac{3\cdot 2^{18} \pi^6h^2 \gamma^2 M_\eta^2}{35(\lambda'^2+3h^2)^3}\\
&-Ht_1\frac{h^2}{16\pi^3M_\eta^2\gamma^3}\biggg[\frac { 3M_\eta \gamma } { 8} \left( 1 + \log \frac { M_\eta } { M_\chi } \right)+\frac{M_\eta^8}{(M_\eta^4+4M_\chi^2/\gamma^2)^2}-\frac{M_\eta^4}{M_\eta^4+4M_\chi^2/\gamma^2}\\
&-\frac{4M_\chi^2M_\eta^4/\gamma^2}{(M_\eta^4+4M_\chi^2/\gamma^2)^2}+\frac{8M_\chi^4/\gamma^2}{M_\eta^2(M_\eta^4+4M_\chi^2/\gamma^2)}+\log\left(\frac{1}{4}\right)-1+\log\left(\frac{M_\eta^6}{M_\chi^4\sqrt{M_\eta^4+4M_\chi^2/\gamma^2}}\right)\biggg]\\
&\hphantom{+}
+\biggg[\left(\frac{32\pi\gamma h^2}{\lambda'^2+3h^2}+\frac{32\pi\gamma \lambda^2}{\lambda^2+3h^2}\right)\left(1+\log\frac{M_\chi\gamma}{2}+\frac{4}{3}Ht_1\right)\\
&+H\left(\frac{2^{13}\pi^4\gamma^2 h^2}{\left(\lambda'^2+3h^2\right)^2}+\frac{2^{13}\pi^4\gamma^2 \lambda^2}{\left(\lambda^2+3h^2\right)^2}-\frac{3\lambda^2}{256\pi\gamma M_\eta^3}-\frac{3h^2}{256\pi\gamma M_\chi^3}\right)\biggg]\frac{\varphi(t_1)^2}{3}\biggg\}\,\dot{\varphi}(t_1) =0~.
\end{aligned}
\end{equation}
\section{Conclusions and Discussions}\label{sec:discussion}

In this paper, we considered a model of two scalar fields $\phi$ and $\chi$ with quartic coupling~(eq.~\ref{eqn:action}). The mass of the $\phi$ field is assumed to be larger than twice the mass of the background field, $\chi$,  so that the  $\phi$ particles can decay into the $\chi$ particles  for  the studying of the  dissipation effects.               
In a thermal bath made up of the $\chi$ particles,  we studied the dynamics of the thermal average of the scalar field       $\phi$ in an  expanding or a contracting universe.  We have assumed the background to be de Sitter spacetime with Hubble parameter $H\neq 0$.  The  Hubble parameter  is taken to be  a small constant in our calculations,  our  results are applicable to other expanding or contracting cosmic backgrounds  to first order in $H$. 
The effective temperature is  taken to be extremely high in our  analysis to have the most manageable configuration yet  retaining  the interesting physics.

From the effective action of $\phi$, we obtained its equation of motion~(eq.~\ref{eqn: effective_EoM}),  which is determined by the thermally  and quantum-mechanically  corrected self-energy and  the self-coupling,  and their derivatives.  The analytic expressions  of these quantities~(eq.~\ref{eqn:kappa}), (eq.\ref{eqn:kappatilde}), (eq.~\ref{kappa_tilde_first_derivative}), (eq.~\ref{sunset_result}) and (eq.~\ref{kappa_tilde_second_derivative}) 
are the main results of this paper.  In the computation of these quantities we expanded the propagator to first order in $H$; and our  results match with  those in flat space, using Minkowski-space propagators in loop diagrams~\cite{Cheung:2015iqa},  if we set $H$ to zero.  Also, we developed some mathematical techniques when calculating these relevant  Feynman diagrams.

In the equation of motion~(eq.~\ref{eqn: effective_EoM}), since the derivatives of $ \kappa $ and $ \tilde{\kappa} $ are derived from 1-loop corrected propagators, they appear to be of higher orders than $ \kappa$ and $\tilde{\kappa} $,  which are   ignored  in the effective potential of $ \phi $.  However, as shown in~(eq.~\ref{kappa_tilde_first_derivative}) and~(eq.~\ref{sunset_result}),  there are terms that do not tend to zero as the coupling constants go to zero. Therefore they cannot  be  excluded from  the effective potential.

In our calculations we have used the assumption that the Hubble parameter $H$ is small enough such that we can expand all terms of interest  to  first order in $H$. Thus,  strictly speaking,  the background  universe is not a purely  de Sitter spacetime, and the cosmic expansion is not strictly exponential.  However, in such a situation we can still obtain  valuable  information about the effects of cosmic expansion/contraction  on the scalar field dynamics.  For example, from equation~(\ref{sunset_result}), which contributes to the dissipation coefficient,  we  find  that a de Sitter space,  which is very close to a flat spacetime,  has the  chance of showing curved-spacetime features comparable to the flat spacetime features if the coupling constants $ h $ and $\lambda'$ are small enough.  This is because in a de Sitter spacetime, we have to  integrate over resonances due to the lack of spacetime translation invariance.

Our assumptions  require   that the temperature be much higher than the masses of the particles and the scale of   the Hubble parameter.  The effective temperature  decreases  as the universe expands,  the corresponding  approximation will fail  when the temperature is of the same order as the masses.   This happens only  near the end of reheating; and thus the working  assumptions are valid for the entire analysis  if our results are applied to  the reheating dynamics.      
Let us stress that  our results also apply to a negative Hubble parameter $H$.  We shall be using these results to  study the quantum dissipative effects in  the  process of matter creation in the CST bounce universe~\cite{Li:2011nj}.

To discuss the quantum dissipative effects in the extreme conditions  in the early universe,  a rigorous theoretical framework of first-principle high temperature thermal quantum field theory is needed. 
The result we have obtained gives us a sense of the difference between the behaviour of a scalar field in a flat spacetime and in a de Sitter spacetime, which is helpful to the study of more realistic and more complicated situations, e.g. the effects of the expansion of the universe on the thermal damping rates of particles in the early universe and the production of matter within some specific inflationary or bounce models.

In this paper, we only calculated the results to the first order in $H$. Theoretically our approach can be extended to arbitrarily high orders, but such attempts are not practical as the integrals get much more complicated. Therefore, when $H$ becomes large enough, for which the approximation we have used would fail, one needs to seek other methods to extract the interesting physics besides matter productions.  

\section*{Acknowledgments}

This research project has been supported in parts by the NSF China
under Contract No.~11775110, and No.~11690034.
We also acknowledge the European Union's Horizon 2020 research and innovation programme (RISE) under the Marie Sk\'lodowska-Curie  grant agreement No.~644121, and  the Priority Academic Program Development for Jiangsu Higher Education Institutions (PAPD).
We would like to express our gratitude to  Jin U Kang and Marco Drewes for useful discussions  and comments on the manuscript.   
L.M. and H.X. thank  Ella Yang for useful suggestions  on  improving their  draft.	
	
\appendix
\section{Free Spectral Function of a Scalar Field}   
\label{appendix:free-spectral-function}
In this appendix  we calculate the free spectral function of a scalar field $\xi$ in de Sitter spacetime with action
\begin{equation}
	\begin{aligned}
		S&=-\int d^4x\sqrt{-g}\,\,\frac{1}{2}\xi\left[\frac{1}{\sqrt{-g}}\partial _\mu\left(\sqrt{-g}g^{\mu\nu}\partial _\nu\right)+m_{\xi}^{2}\right]\xi\\
		&=-\int d^4x\,\,\frac{a^3}{2}\xi\left[\left(3H\partial _t+\partial _{t}^{2}-a^{-2}\nabla^2\right)+m_{\xi}^{2}\right]\xi,
	\end{aligned}
\end{equation}
where   $a=e^{Ht} $.

If we set $ \tau=-e^{-Ht}/H $, $ \tilde{\xi}=a(t)\xi(x) $, then the action becomes
\begin{equation}
	S=-\int d\tau\int d^3x\frac{1}{2}\tilde{\xi}\left[\partial _{\tau}^{2}-\nabla^2+\left(m_\xi^2-2H^2\right)a^2\right]\tilde{\xi}~, 
\end{equation}
from which we obtain the equation of motion for $ \tilde{\xi} $:
\begin{equation}
	\left[\frac{\partial ^2}{\partial \tau^2}-\nabla^2+(m_\xi^2-2H^2)a^2\right]\tilde{\xi}(\boldsymbol{x},\tau)=0.
\end{equation}
If  we expand $ \tilde{\xi}(\boldsymbol{x},\tau) $ as
\begin{equation}
	\tilde{\xi}(\boldsymbol{x},\tau)=\int \frac{d^3k}{(2\pi)^3}\tilde{\xi}(\boldsymbol{k},\tau)e^{i\boldsymbol{k}\cdot \boldsymbol{x}}+h.c.~.
\end{equation}

To  solve the equation of motion we employ  the WKB method, and obtain, 
\begin{equation}\label{xi-expansion}
	\tilde{\xi}(\boldsymbol{x},\tau)=\int \frac{d^3k}{(2\pi)^3}\left\{
	A_k \frac{\exp\left[\displaystyle -i\int_{\tau_0}^{\tau}\omega(\tau')d\tau'\right]}{\sqrt{2\omega(\tau)}}e^{i\boldsymbol{k}\cdot \boldsymbol{x}}+h.c.~, 
	\right\},
\end{equation}
where $ \tau_0 $ is a constant of integration,  $ A_k $  a momentum-dependent operator, and 
\begin{equation}
	\omega(\tau)=\sqrt{\boldsymbol{k}^2+(m_\xi^2-2H^2)a^2}~. 
\end{equation}

The conjugate momentum is given by, 
\begin{equation}
	\tilde{\pi}(\boldsymbol{x},\tau)=\int \frac{d^3k}{(2\pi)^3}\frac{1}{\sqrt{2}}\left[-\frac{1}{2}H(m_\xi^2-2H^2)a^3\omega^{-5/2}-i\omega^{1/2}\right]A_k\exp\left(-i\int_{\tau_0}^{\tau}\omega(\tau')d\tau'\right)e^{i\boldsymbol{k}\cdot \boldsymbol{x}}+h.c.~. 
\end{equation}

From the commutation relation $ \left[\tilde{\xi}(\boldsymbol{x},\tau),\tilde{\pi}(\boldsymbol{x}',\tau)\right] =i \delta^3(\boldsymbol{x}-\boldsymbol{x}')$ we obtain $ \left[A_{\boldsymbol{k}},A_{\boldsymbol{k}'}^\dagger\right]=\delta^3(\boldsymbol{k}-\boldsymbol{k}') $. With this commutation relation of 
$ A_{\boldsymbol{k}}$ and $A_{\boldsymbol{k}'}^\dagger $ we can calculate the free spectral function 
using~(\ref{xi-expansion}):
\begin{equation}
	\begin{aligned}
		\Delta^-(\boldsymbol{k},t_1,t_2)&=i\left\langle\left[\frac{\tilde{\xi}(\tau_1)}{a(t_1)},\frac{\tilde{\xi}(\tau_2)}{a(t_2)}\right]\right\rangle_\beta\\
		&=\frac{1}{a(t_1)^{3/2}a(t_2)^{3/2}\sqrt{\Omega_\xi(t_1)}\sqrt{\Omega_\xi(t_2)}}\sin\left[\int_{t_2}^{t_1}\Omega_\xi(t')dt'\right],
	\end{aligned}
\end{equation}
where
\begin{equation}
	\Omega_\xi(t)\equiv \sqrt{\boldsymbol{k}^2/a(t)^2+(m_\xi^2-2H^2)}~. 
\end{equation}

\section{A few  useful formulae}

In this appendix  we present  the  relevant formulae that we  have used in computing the integrals 
associated with  the Feynman diagrams. 

\subsection{Integrals Involving the Bose Distribution function}
\label{appendix: bose distribution integral}
When doing the integrals in~(eq.~\ref{eqn:tadpole}), we  encounter expressions of the following form,
\begin{equation}\label{eqn:h}
	h_n(y)=\frac{1}{\Gamma(n)}\int_{0}^{\infty} dx\frac{x^{n-1}}{\sqrt{x^2+y^2}}\frac{1}{e^{\sqrt{x^2+y^2}}-1}\quad (n\in \mathbb{Z}^{+}).
\end{equation}
By expanding $ 1/\{[\exp(\sqrt{x^2+y^2})]-1\} $ around $ \sqrt{x^2+y^2}=0 $ into a Laurent series, 
 integrating over    $ x $ and then expanding the expression around $ y=0 $, we obtain
\begin{equation}
	\begin{aligned}
		h_1(y)=\frac{\pi}{2y}+\frac{1}{2}\log\frac{y}{4\pi}+\frac{\gamma_E}{2}+\sum_{m=1}^{\infty}\frac{(-1)^{m}}{2^{m+1}}\frac{(2m-1)!!}{m!}\zeta(2m+1)\left(\frac{y}{2\pi}\right)^{2m}~.
	\end{aligned}
\end{equation}
It is easy to check that $ h_n(y) $ satisfies,
\begin{equation}
	\frac{dh_{n+1}}{dy}=-\frac{yh_{n-1}}{n}~.
\end{equation}
Setting $ n=2 $ and integrating the above equality we have,
\begin{equation}\label{eqn: h_3}
	\begin{aligned}
		h_3(y)&=-\frac{1}{2}\int yh_1(y)dy+\frac{\pi^2}{12}\\
		&=\frac{\pi^2}{12}-\frac{\pi y}{4}-\left(\frac{\gamma_E}{8}-\frac{1}{16}\right)y^2-\frac{y^2}{8}\log\frac{y}{4\pi}+\sum_{m=1}^{\infty}\frac{(-1)^{m+1}}{2^{m+3}}\frac{(2m-1)!!}{(m+1)!}\frac{\zeta(2m+1)}{(2\pi)^{2m}}y^{2m+2}.
	\end{aligned}
\end{equation}

\subsection{Fourier Transformation of a General Form}
\label{appendix: Fourier of general form}
In this subsection we shall  calculate the real and imaginary parts of the following expressions:
\begin{equation}
	\begin{aligned}
		&F^\omega\left[\theta(t)\mathrm{Im}\int \frac{d^3q}{(2\pi)^3}\left(K_0+t K_1+t^2K_2\right)e^{-2i\omega_q t-\alpha t/\omega_q}\right],\\
		&F^\omega\left[\theta(t)\mathrm{Im}\int \frac{d^3q}{(2\pi)^3}\left(K_0+t K_1+t^2K_2\right)e^{-\alpha t/\omega_q}\right],
	\end{aligned}
\end{equation}
where $ F^\omega $ denotes a Fourier transformation   w.r.t.  $ t $ and $ K_0, K_1,K_2 $ are rational functions of 
$ \boldsymbol{q}^2 $.
We  use $ K $ to denote any of $ K_0,K_1,K_2 $ and calculate the following integral:
\begin{equation}
	\begin{aligned}
		&F^{\omega}\left[ \theta(t) \operatorname{Im} \int \frac{d^{3} q}{(2 \pi)^{3}} K\left(\omega_{q}\right) e^{-2 i \omega_{q} t- \frac{\alpha}{\omega_q} t}\right]\\
		=& \int \frac{d^3 q}{(2\pi)^3}\left[-i\mathrm{Re}K(\omega_q)F^\omega\left(\theta(t)\frac{e^{-2i\omega_q t}-e^{2i\omega_q t}}{2}e^{-\frac{\alpha}{\omega_q}t}\right)+\mathrm{Im}K(\omega_q)F^\omega\left(\theta(t)\frac{e^{-2i\omega_q t}+e^{2i\omega_q t}}{2}e^{-\frac{\alpha}{\omega_q}t}\right)\right].
	\end{aligned}
\end{equation}
Carrying out the Fourier transformation  in the integrand and change the variable of integration to $ \omega_q=\sqrt{\boldsymbol{q}^2+m^2} $, the integral  becomes,
\begin{equation}
	\begin{aligned}
		&\frac{1}{(2 \pi)^{3}} \int_{m}^{\infty} d \omega_{q} \frac{\omega_{q}}{\sqrt{\omega_{q}^{2}-m^{2}}} 4 \pi\left(\omega_{q}^{2}-m^{2}\right)\\
		&\quad\times \bigg\{ \mathrm{Re}K(\omega_q)\left[\frac{1}{2}\left(\dfrac{\omega-2\omega_q}{(\omega-2\omega_q)^2+(\alpha/\omega_q)^2}-\frac{\omega+2\omega_q}{(\omega+2\omega_q)^2+(\alpha/\omega_q)^2}\right)\right.\\
		&\qquad\quad \left.-i\left(\frac{\alpha/(2\omega_q)}{(\omega-2\omega_q)^2+(\alpha/\omega_q)^2}-\frac{\alpha/(2\omega_q)}{(\omega+2\omega_q)^2+(\alpha/\omega_q)^2}\right)\right]\\
		&\qquad+  \mathrm{Im}K(\omega_q)\left[\frac{i}{2}\left(\dfrac{\omega+2\omega_q}{(\omega+2\omega_q)^2+(\alpha/\omega_q)^2}+\frac{\omega-2\omega_q}{(\omega-2\omega_q)^2+(\alpha/\omega_q)^2}\right)\right.\\
		&\qquad\quad \left.+\left(\frac{\alpha/(2\omega_q)}{(\omega+2\omega_q)^2+(\alpha/\omega_q)^2}+\frac{\alpha/(2\omega_q)}{(\omega-2\omega_q)^2+(\alpha/(2\omega_q))^2}\right)\right]\bigg\}.
	\end{aligned}
\end{equation}
Since we only calculate to the lowest order in $ \alpha $, the above expression  can be further simplified to,
\begin{equation}
	\begin{aligned}
		&\frac{1}{(2 \pi)^{3}} \int_{m}^{\infty} d \omega_{q} \frac{\omega_{q}}{\sqrt{\omega_{q}^{2}-m^{2}}} 4 \pi\left(\omega_{q}^{2}-m^{2}\right)\\
		&\quad \times\left\{\mathrm{Re}K(\omega_q)\left[\frac{2\omega_q}{\omega^2-4\omega_q^2}-i\left(\frac{\alpha}{2\omega_q(\omega-2\omega_q)^2}-\frac{\alpha}{2\omega_q(\omega+2\omega_q)^2}\right)\right]\right.\\
		&\qquad\left.+\mathrm{Im}K(\omega_q)\left[i\frac{\omega}{\omega^2-4\omega_q^2}+\left(\frac{\alpha}{2\omega_q(\omega-2\omega_q)^2}+\frac{\alpha}{2\omega_q(\omega+2\omega_q)^2}\right)\right]\right\}\\
		&\equiv I_{re}\left[K(\omega_q)\right]+iI_{im}\left[K(\omega_q)\right],
	\end{aligned}
\end{equation}
where
\begin{equation}\left\{
	\begin{aligned}
		I_{r e}\left[K\left(\omega_{{q}}\right)\right]=&\frac{1}{(2 \pi)^{3}} \int_{m}^{\infty} d \omega_{{q}} 4 \pi \omega_{{q}} \sqrt{\omega_{{q}}^{2}-m^2}\Bigg\{\operatorname{Re} K\left(\omega_{{q}}\right) \frac{2 \omega_{{q}}}{\omega^{2}-4 \omega_{{q}}^{2}}\\
		&+\operatorname{Im} K\left(\omega_{{q}}\right)\left[\frac{\alpha}{2\omega_{{q}}\left(\omega+2 \omega_{{q}}\right)^{2}}+\frac{\alpha}{2\omega_{{q}}\left(\omega-2 \omega_{{q}}\right)^{2}}\right]\Bigg\}\\
		I_{im}\left[K\left(\omega_{{q}}\right)\right]=&\frac{1}{(2 \pi)^{3}} \int_{m}^{\infty} d \omega_{{q}} 4 \pi \omega_{{q}} \sqrt{\omega_{{q}}^{2}-m^2}\Bigg\{\operatorname{Im} K\left(\omega_{{q}}\right) \frac{\omega}{\omega^{2}-4 \omega_{{q}}^{2}}\\
		&-\operatorname{Re} K\left(\omega_{{q}}\right)\left[\frac{\alpha}{2\omega_{{q}}\left(\omega-2 \omega_{{q}}\right)^{2}}-\frac{\alpha}{2\omega_{{q}}\left(\omega+2 \omega_{{q}}\right)^{2}}\right]\Bigg\}
	\end{aligned}\right.~.
\end{equation}
We arrive at,  
\begin{equation}\label{eqn: fourier_Im}
	\begin{aligned}
		& \operatorname{Im} F^{\omega}\left[\theta(t) \operatorname{Im} \int \frac{d^{3} q}{(2 \pi)^{3}}\left(K_{0}+t K_{1}+t^{2} K_{2}\right) e^{-2 i \omega_{q} t- \frac{\alpha}{\omega_{q}} t}\right]\\
		=&\operatorname{Im} F^{\omega} \left[\theta(t) \operatorname{Im} \int \frac{d^{3} q}{(2 \pi)^{3}}K_{0} e^{-2 i \omega_{q} t- \frac{\alpha}{\omega_{q}} t}\right]+\operatorname{Im}\Bigg\{\left(-i \frac{\partial}{\partial \omega}\right) F^{\omega} \left[\theta(t) \operatorname{Im} \int \frac{d^{3} q}{(2 \pi)^{3}}K_{1}e^{-2 i \omega_{q} t-\frac{\alpha}{\omega_{q}}t} \right]\Bigg\}\\
		& +\operatorname{Im}\Bigg\{\left(-\frac{\partial^{2}}{\partial \omega^{2}}\right) F^{\omega} \left[\theta(t) \operatorname{Im} \int \frac{d^{3} q}{(2 \pi)^{3}}K_{3}e^{-2 i \omega_{q} t-\frac{\alpha}{\omega_{q}} t}\right]\Bigg\}\\
	=& I_{i m}\left[K_{0}\right]-\frac{\partial}{\partial \omega} I_{r e}\left[K_{1}\right]-\frac{\partial^{2}}{\partial \omega^{2}} I_{i m}\left[K_{2}\right]~.
	\end{aligned}
\end{equation}
In a similar way  we obtain,
\begin{equation}\label{eqn: fourier_Re}
	\operatorname{Re} F^{\omega}\left[\theta(t) \operatorname{Im} \int \frac{d^{3} q}{(2 \pi)^{3}}\left(K_{0}+t K_{1}+t^{2} K_{2}\right) e^{-2 i \omega_{q} t- \frac{\alpha}{\omega_{q}} t}\right]=I_{re}\left[K_{0}\right]+\frac{\partial}{\partial \omega} I_{im}\left[K_{1}\right]-\frac{\partial^{2}}{\partial \omega^{2}} I_{re}\left[K_{2}\right]~,
\end{equation}
and
\begin{equation}\label{eqn: fourier_cross}
	\begin{aligned}
		&F^{\omega}\left[\theta(t) \operatorname{Im} \int \frac{d^{3} q}{(2 \pi)^{3}}\left(K_{0}+t K_{1}+t^{2} K_{2}\right) e^{- \frac{\alpha}{\omega_{q}} t}\right]\\
		=&\int \frac{d^3q}{(2\pi)^3}\left(\mathrm{Im}K_0\frac{\omega_q \alpha}{\alpha^2+\omega^2\omega^2_q}+\mathrm{Im}K_1\frac{\partial}{\partial \omega}\frac{\omega\omega^2_q}{\alpha^2+\omega^2\omega^2_q}-\mathrm{Im}K_2\frac{\partial^2}{\partial \omega^2}\frac{\omega_q \alpha}{\alpha^2+\omega^2\omega^2_q}\right)\\
		&+i\int \frac{d^3q}{(2\pi)^3}\left(\mathrm{Im}K_0\frac{\omega\omega^2_q}{\alpha^2+\omega^2\omega^2_q}-\mathrm{Im}K_1\frac{\partial }{\partial \omega}\frac{\omega_q\alpha}{\alpha^2+\omega^2\omega^2_q}-\mathrm{Im}K_2\frac{\partial ^2}{\partial \omega^2}\frac{\omega\omega^2_q}{\alpha^2+\omega^2\omega^2_q}\right)~.
	\end{aligned}
\end{equation}

\section{Calculation of the Tadpole Diagram}
In this appendix  we present  the detailed  calculation of the the tadpole diagram.  
Inserting the propagator~(eq.~\ref{propagator+-}) into~(eq.~\ref{eqn:tadpole}) we have,
\begin{equation}
	\begin{aligned}
		&\Pi^R(t_1,t_2)\\
		=&\frac{1}{2}\delta(t_1-t_2)\sum_{\xi=\eta,\chi}\int \frac{d^3p}{(2\pi)^3}\frac{g_\xi}{a(t_1)^3\sqrt{\boldsymbol{p}^2/a(t_1)^2+M^2_\xi}}\left[\frac{1}{2}+\frac{1}{e^{\sqrt{\boldsymbol{p}^2\beta(t_1)^2/a(t_1)^2+M^2_\xi\beta(t_1)^2}}-1}\right]\\
		=&\delta(t_1-t_2)\frac{1}{4\pi^2\beta(t_1)^2}\sum_{\xi=\eta,\chi}\int dx\,g_\xi\left[\frac{1}{2}\frac{x^2}{\sqrt{x^2+M^2_\xi\beta(t_1)^2}}+\frac{x^2}{\sqrt{x^2+M^2_\xi\beta(t_1)^2}}\frac{1}{e^{\sqrt{x^2+M_\xi^2\beta(t_1)^2}}-1}\right]~.
	\end{aligned}
\end{equation}
We  replace the divergent integrals that are independent of time by the constants $ C_1 $ and $ C_2 $:
\begin{equation}
	\begin{aligned}
		&C_1=\frac{1}{2}\sum_{\xi=\eta,\chi}\int \frac{d^3p}{(2\pi)^3}\frac{g_\xi}{2\sqrt{\boldsymbol{p}^2+M_\xi^2}}+\text{counter terms},\\
		&C_2=\frac{1}{2}\sum_{\xi=\eta,\chi}\int\frac{d^3p}{(2\pi)^3}\left[\frac{g_\xi}{\sqrt{\boldsymbol{p}^2+M_\xi^2}}+\frac{g_\xi M_\xi^2}{2(\boldsymbol{p}^2+M_\xi^2)^{3/2}}\right]+\text{counter terms.}
	\end{aligned}
\end{equation}
yielding   $\Pi^R(t_1,t_2)$ in the following form, 
\begin{equation}
	\Pi^R(t_1,t_2)=\delta(t_1-t_2)\left\{C_1-C_2Ht_1+\frac{1}{2\pi^2\beta(t_1)^2}\sum_{\xi=\eta,\chi}g_\xi h_3\left[M_\xi\beta(t_1)\right]\right\}~.
\end{equation}
Comparing with  the case of  $ H=0 $,
\begin{equation}
	\Pi^R(t_1,t_2)\approx\delta(t_1-t_2)\left[C_1+ \sum_{\xi=\eta,\chi}\int\frac{d^3p}{(2\pi)^3}\frac{g_\xi}{2\sqrt{\boldsymbol{p}^2+M_\xi^2}}\cdot\frac{1}{\exp\left(\dfrac{\beta_0\sqrt{\boldsymbol{p}^2+M_\xi^2}}{a_0}\right)-1}\right],
\end{equation}
we  can conclude  that $ C_1 =0 $ at zero temperature in flat spacetime.  
Furthermore the propagator of $ \phi $ has a pole at the physical mass: the first term and the second term in the $ C_2 $ integral are proportional to $ C_1 $ or to the derivative of $ C_1 $ with respect to $ M_\xi $,  
we  deduce that  $ C_2=0 $.  The renormalised  $\Pi^R(t_1,t_2)$ is finally given by, 
\begin{equation}
	\Pi^R(t_1,t_2)=\delta(t_1-t_2)\left\{\frac{1}{2\pi^2\beta(t_1)^2}\sum_{\xi=\eta,\chi}g_\xi h_3\left[M_\xi\beta(t_1)\right]\right\}~.
\end{equation}

\section{Intermediate Results in the Calculation of the Sunset Diagram}
\label{appendix: C3}
Let us now turn to the  contribution of the sunset diagram to the self-energy, 
\begin{equation}
	\begin{aligned} 
		&-i\left[\Pi^R(t_1,t_2)\right]_{\text{sunset}}\\
		=&-ih^2\theta(t_1-t_2)\int \frac{d^3k}{(2\pi)^3}\frac{d^3l}{(2\pi^3)}\mathrm{Im}\left[(\Delta_\chi)_>(\boldsymbol{k},t_1,t_2)(\Delta_\chi)_>(\boldsymbol{l},t_1,t_2)(\Delta_\eta)_>(\boldsymbol{k}+\boldsymbol{l},t_1,t_2)\right]~.
	\end{aligned}
\end{equation}
By expanding the propagators to  first order in $ H $ and inserting them into the above equation we obtain the following expression of the imaginary part of $ \kappa $,
\begin{equation}\label{eqn:kappa_H_expansion}  
	\begin{aligned}
		&\mathrm{Im}\kappa(t_1,\omega)=\frac{h^2}{8} \mathrm{Im}F^\omega\Bigg\{\theta(t)\int \frac{d^{3}k}{(2\pi)^3}\frac{d^3l}{(2\pi)^3}\\
		&\mathrm{Im}\frac { 1 } { \omega _ { \eta q } \omega_{\chi k} \omega_{\chi l} }\left[\left( 1 - 9 H t _ { 1 } + \frac { 9 } { 2 } H t \right) +\left( H t _ { 1 } - \frac { 1 } { 2 } H t \right) \left(3 - \frac { M_\eta ^ { 2 } } { \omega _ { \eta q } ^ { 2 } } - \frac { M_\chi ^ { 2 } } { \omega_{\chi k} ^ { 2 } } - \frac { M_\chi^ { 2 } } { \omega_{\chi l} ^ { 2 } } \right)\right.\\
		&\left. - H \left( t _ { 1 } - t \right)  \left( f (\omega_{\eta q}) \frac { \gamma M_\eta ^ { 2 } } { \omega _ { \eta q } } + f (\omega_{\chi k}) \frac { \gamma M_\chi^ { 2 } } { \omega_{\chi k} } + f (\omega_{\chi l}) \frac { \gamma M_\chi ^ { 2 } } { \omega_{\chi l} } \right) + i H \left( 2 t _ { 1 } t - t ^ { 2 } \right) \left( \frac { \boldsymbol { q } ^ { 2 } } { 2 \omega _ { \eta q } } + \frac { \boldsymbol { k } ^ { 2 } } { 2 \omega_{\chi k} } + \frac { \boldsymbol { l } ^ { 2 } } { 2 \omega_{\chi l} } \right)\right]\\
		&\times \left( 1 + f (\omega_{\eta q})\right) \left( 1 + f (\omega_{\chi k}) \right) \left( 1 + f (\omega_{\chi l}) \right) \exp \left[ - i \left( \omega _ { \eta q } + \omega_{\chi k} + \omega_{\chi l} \right) t-\Gamma_\chi(\omega_{\chi k})t/2-\Gamma_\chi(\omega_{\chi l})t/2 \right]\Bigg\} \\
		&+ \left( \omega _ { \eta q } \rightarrow - \omega _ { \eta q } , \omega_{\chi k} \rightarrow - \omega_{\chi k} , \omega_{\chi l} \rightarrow - \omega_{\chi l} \right)\\
		& + \left( \omega _ { \eta q } \rightarrow - \omega _ { \eta q } , \omega_{\chi k} \rightarrow - \omega_{\chi k} , \omega_{\chi l} \rightarrow \omega_{\chi l} \right) + \left( \omega _ { \eta q } \rightarrow \omega _ { \eta q } , \omega_{\chi k} \rightarrow \omega_{\chi k} , \omega_{\chi l} \rightarrow - \omega_{\chi l} \right)\\
		& + \left( \omega _ { \eta q } \rightarrow - \omega _ { \eta q } , \omega_{\chi k} \rightarrow \omega_{\chi k} , \omega_{\chi l} \rightarrow - \omega_{\chi l} \right) + \left( \omega _ { \eta q } \rightarrow \omega _ { \eta q } , \omega_{\chi k} \rightarrow - \omega_{\chi k} , \omega_{\chi l} \rightarrow \omega_{\chi l} \right)\\
		& + \left( \omega _ { \eta q } \rightarrow \omega _ { \eta q } , \omega_{\chi k} \rightarrow - \omega_{\chi k} , \omega_{\chi l} \rightarrow - \omega_{\chi l} \right) + \left( \omega _ { \eta q } \rightarrow - \omega _ { \eta q } , \omega_{\chi k} \rightarrow \omega_{\chi k} , \omega_{\chi l} \rightarrow \omega_{\chi l} \right)\\
		&=\frac{h^2}{8} \mathrm{Im}F^\omega\Bigg\{\theta(t)\int \frac{d^{3}k}{(2\pi)^3}\frac{d^3l}{(2\pi)^3}\frac { 1 } { \omega _ { \eta q } \omega_{\chi k} \omega_{\chi l} }\Bigg\{-\sin \left[ \left( \omega _ { \eta q } + \omega_{\chi k} + \omega_{\chi l} \right) t \right]\exp \left(-\Gamma_\chi(\omega_{\chi k})t/2-\Gamma_\chi(\omega_{\chi l})t/2\right) \times\\
		&\left[ \left( 1 - 9 H t _ { 1 } \right) \gamma _ { 0 } \left( \omega _ { \eta q } , \omega_{\chi k} , \omega_{\chi l} \right) + H t _ { 1 } \gamma _ { 1 } \left( \omega _ { \eta q } , \omega_{\chi k} , \omega_{\chi l} \right) - H t _ { 1 } \gamma _ { 2 } \left( \omega _ { \eta q } , \omega_{\chi k} , \omega_{\chi l} \right) \right] \\
		&+t \cos \left[ \left( \omega _ { \eta q } + \omega_{\chi k} + \omega_{\chi l} \right) t \right]\exp \left(-\Gamma_\chi(\omega_{\chi k})t/2-\Gamma_\chi(\omega_{\chi l})t/2\right)\cdot2 H t _ { 1 } \gamma _ { 3 } \left( \omega _ { \eta q } , \omega_{\chi k} , \omega_{\chi l} \right) \\
		&-t \sin \left[ \left( \omega _ { \eta q } + \omega_{\chi k} + \omega_{\chi l} \right) t \right]\exp \left(-\Gamma_\chi(\omega_{\chi k})t/2-\Gamma_\chi(\omega_{\chi l})t/2\right)\\
		&\times \Bigg[ \frac { 9 } { 2 } H \gamma _ { 0 } \left( \omega _ { \eta q } , \omega_{\chi k} , \omega_{\chi l} \right)- \frac { 1 } { 2 } H \gamma _ { 1 } \left( \omega _ { \eta q } , \omega_{\chi k} , \omega_{\chi l} \right) + H \gamma _ { 2 } \left( \omega _ { \eta q } , \omega_{\chi k} , \omega_{\chi l} \right) \Bigg] \\
		&- t ^ { 2 } \cos \left[ \left( \omega _ { \eta q } + \omega_{\chi k} + \omega_{\chi l} \right) t \right]\exp \left(-\Gamma_\chi(\omega_{\chi k})t/2-\Gamma_\chi(\omega_{\chi l})t/2\right) \cdot H \gamma _ { 3 } \left( \omega _ { \eta q } , \omega_{\chi k} , \omega_{\chi l} \right)\Bigg\}\\
		&\times \left( 1 + f (\omega_{\eta q}) \right) \left( 1 + f(\omega_{\chi k}) \right) \left( 1 + f(\omega_{\chi l}) \right)\Bigg\} \\
		&+ \left( \omega _ { \eta q } \rightarrow - \omega _ { \eta q } , \omega_{\chi k} \rightarrow - \omega_{\chi k} , \omega_{\chi l} \rightarrow - \omega_{\chi l} \right)\\
		& + \left( \omega _ { \eta q } \rightarrow - \omega _ { \eta q } , \omega_{\chi k} \rightarrow - \omega_{\chi k} , \omega_{\chi l} \rightarrow \omega_{\chi l} \right) + \left( \omega _ { \eta q } \rightarrow \omega _ { \eta q } , \omega_{\chi k} \rightarrow \omega_{\chi k} , \omega_{\chi l} \rightarrow - \omega_{\chi l} \right)\\
		& + \left( \omega _ { \eta q } \rightarrow - \omega _ { \eta q } , \omega_{\chi k} \rightarrow \omega_{\chi k} , \omega_{\chi l} \rightarrow - \omega_{\chi l} \right) + \left( \omega _ { \eta q } \rightarrow \omega _ { \eta q } , \omega_{\chi k} \rightarrow - \omega_{\chi k} , \omega_{\chi l} \rightarrow \omega_{\chi l} \right)\\
		& + \left( \omega _ { \eta q } \rightarrow \omega _ { \eta q } , \omega_{\chi k} \rightarrow - \omega_{\chi k} , \omega_{\chi l} \rightarrow - \omega_{\chi l} \right) + \left( \omega _ { \eta q } \rightarrow - \omega _ { \eta q } , \omega_{\chi k} \rightarrow \omega_{\chi k} , \omega_{\chi l} \rightarrow \omega_{\chi l} \right),
	\end{aligned}
\end{equation}
where $ \Gamma_\chi(\omega_{\chi k})\approx\frac{\lambda'^2+3h^2}{256\pi^3\gamma^2\omega_{\chi k}}, \Gamma_\chi(\omega_{\chi l})\approx\frac{\lambda'^2+3h^2}{256\pi^3\gamma^2\omega_{\chi l}} $~\cite{Cheung:2015iqa} and
\begin{equation}\left\{
	\begin{aligned}
		&\gamma _ { 0 } \left( \omega _ { \eta q } , \omega_{\chi k} , \omega_{\chi l} \right) = 1~,\\
		&\gamma _ { 1 } \left( \omega _ { \eta q } , \omega_{\chi k} , \omega_{\chi l} \right) = 3 - \frac { M_\eta ^ { 2 } } { \omega _ { \eta q } ^ { 2 } } - \frac { M_\chi ^ { 2 } } { \omega_{\chi k} ^ { 2 } } - \frac { M_\chi^ { 2 } } { \omega_{\chi l} ^ { 2 } }~,\\
		&\gamma _ { 2 } \left( \omega _ { \eta q } , \omega_{\chi k} , \omega_{\chi l} \right) = f (\omega_{\eta q}) \frac { \gamma M_\eta ^ { 2 } } { \omega _ { \eta q } } + f (\omega_{\chi k}) \frac { \gamma M_\chi^ { 2 } } { \omega_{\chi k} } + f (\omega_{\chi l})\frac { \gamma M_\chi ^ { 2 } } { \omega_{\chi l} }~,\\
		&\gamma _ { 3 } \left( \omega _ { \eta q } , \omega_{\chi k} , \omega_{\chi l} \right) = \frac { \boldsymbol { q } ^ { 2 } } { 2 \omega _ { \eta q } } + \frac { \boldsymbol { k } ^ { 2 } } { 2 \omega_{\chi k} } + \frac { \boldsymbol { l } ^ { 2 } } { 2 \omega_{\chi l} } ~.
	\end{aligned}\right. 
\end{equation}
We note  that in~(eq.~\ref{eqn:kappa_H_expansion}) there are four kinds of Fourier transformations, 
which involves 
$ \sin(\omega_{\eta q}+\omega_{\chi k}+\omega_{\chi l}), $
$t\cos(\omega_{\eta q}+\omega_{\chi k}+\omega_{\chi l}) , $
 $ t\sin(\omega_{\eta q}+\omega_{\chi k}+\omega_{\chi l})  , $   and 
  $ t^2\cos(\omega_{\eta q}+\omega_{\chi k}+\omega_{\chi l})  , $  respectively.   
For the first two  we carry out the Fourier transformations and perform  the angle integrals and  express them in the following form:
\begin{equation}   
	\begin{aligned}
	 I_s[\gamma_i]  \equiv
   &-\mathrm{Im}F^\omega\Bigg\{\theta(t)\int \frac{d^{3}k}{(2\pi)^3}\frac{d^3l}{(2\pi)^3}\\
  &\sin \left[ \left( \omega _ { \eta q } + \omega_{\chi k} + \omega_{\chi l} \right) t \right]\exp \left(-\Gamma_\chi(\omega_{\chi k})t/2-\Gamma_\chi(\omega_{\chi l})t/2\right) \frac { \gamma_i(\omega_{\eta q},\omega_{\chi k},\omega_{\chi l}) } { \omega _ { \eta q } \omega_{\chi k} \omega_{\chi l} }\\
  &\times\left( 1 + f (\omega_{\eta q}) \right) \left( 1 + f(\omega_{\chi k}) \right) \left( 1 + f(\omega_{\chi l}) \right)\Bigg\}  \\
  + &\left( \omega _ { \eta q } \rightarrow - \omega _ { \eta q } , \omega_{\chi k} \rightarrow - \omega_{\chi k} , \omega_{\chi l} \rightarrow - \omega_{\chi l} \right)\\
  + & \left( \omega _ { \eta q } \rightarrow - \omega _ { \eta q } , \omega_{\chi k} \rightarrow - \omega_{\chi k} , \omega_{\chi l} \rightarrow \omega_{\chi l} \right) + \left( \omega _ { \eta q } \rightarrow \omega _ { \eta q } , \omega_{\chi k} \rightarrow \omega_{\chi k} , \omega_{\chi l} \rightarrow - \omega_{\chi l} \right)\\
   + & \left( \omega _ { \eta q } \rightarrow - \omega _ { \eta q } , \omega_{\chi k} \rightarrow \omega_{\chi k} , \omega_{\chi l} \rightarrow - \omega_{\chi l} \right) + \left( \omega _ { \eta q } \rightarrow \omega _ { \eta q } , \omega_{\chi k} \rightarrow - \omega_{\chi k} , \omega_{\chi l} \rightarrow \omega_{\chi l} \right)\\
+ &\left( \omega _ { \eta q } \rightarrow \omega _ { \eta q } , \omega_{\chi k} \rightarrow - \omega_{\chi k} , \omega_{\chi l} \rightarrow - \omega_{\chi l} \right) + \left( \omega _ { \eta q } \rightarrow - \omega _ { \eta q } , \omega_{\chi k} \rightarrow \omega_{\chi k} , \omega_{\chi l} \rightarrow \omega_{\chi l} \right)\\
	\approx  & \frac { 1 } { 4 ( 2 \pi ) ^ { 3 } } \int _ { M_\chi } ^ { \infty } d \omega_{\chi l} \int _ { M_\chi t _ { 1 } - } ^ { M_\chi t _ { 1 + } } d \omega_{\chi k}\big[\gamma _ { i } \left( - \omega + \omega_{\chi k} + \omega_{\chi l} , - \omega_{\chi k} , - \omega_{\chi l} \right) \left( 1 + f \left( \omega_{\chi k} + \omega_{\chi l} - \omega \right) \right) f \left( \omega_{\chi k} \right) f \left( \omega_{\chi l} \right)\\
		 - &   \gamma _ { i } \left( \omega - \omega_{\chi k} - \omega_{\chi l} , \omega_{\chi k} , \omega_{\chi l} \right) f \left( \omega_{\chi k} + \omega_{\chi l} - \omega \right) \left( 1 + f \left( \omega_{\chi k} \right) \right) \left( 1 + f \left( \omega_{\chi l} \right) \right)\big]-\left(\omega\to -\omega\right)~.
	\end{aligned}
\end{equation}
\begin{equation}     
	\begin{aligned}
		-\frac{\partial}{\partial\omega}I_c[\gamma_i]  \equiv
		&\mathrm{Im}F^\omega\Bigg\{\theta(t)\int \frac{d^{3}k}{(2\pi)^3}\frac{d^3l}{(2\pi)^3}\\
		&t\cos \left[ \left( \omega _ { \eta q } + \omega_{\chi k} + \omega_{\chi l} \right) t \right]\exp \left(-\Gamma_\chi(\omega_{\chi k})t/2-\Gamma_\chi(\omega_{\chi l})t/2\right)\frac { \gamma_i(\omega_{\eta q},\omega_{\chi k},\omega_{\chi l}) } { \omega _ { \eta q } \omega_{\chi k} \omega_{\chi l} }\\
		&\times\left( 1 + f (\omega_{\eta q}) \right) \left( 1 + f (\omega_{\chi k})\right) \left( 1 + f (\omega_{\chi l})\right)\Bigg\}  \\
		&+ \left( \omega _ { \eta q } \rightarrow - \omega _ { \eta q } , \omega_{\chi k} \rightarrow - \omega_{\chi k} , \omega_{\chi l} \rightarrow - \omega_{\chi l} \right)\\
		& + \left( \omega _ { \eta q } \rightarrow - \omega _ { \eta q } , \omega_{\chi k} \rightarrow - \omega_{\chi k} , \omega_{\chi l} \rightarrow \omega_{\chi l} \right) + \left( \omega _ { \eta q } \rightarrow \omega _ { \eta q } , \omega_{\chi k} \rightarrow \omega_{\chi k} , \omega_{\chi l} \rightarrow - \omega_{\chi l} \right)\\
		& + \left( \omega _ { \eta q } \rightarrow - \omega _ { \eta q } , \omega_{\chi k} \rightarrow \omega_{\chi k} , \omega_{\chi l} \rightarrow - \omega_{\chi l} \right) + \left( \omega _ { \eta q } \rightarrow \omega _ { \eta q } , \omega_{\chi k} \rightarrow - \omega_{\chi k} , \omega_{\chi l} \rightarrow \omega_{\chi l} \right)\\
		& + \left( \omega _ { \eta q } \rightarrow \omega _ { \eta q } , \omega_{\chi k} \rightarrow - \omega_{\chi k} , \omega_{\chi l} \rightarrow - \omega_{\chi l} \right) + \left( \omega _ { \eta q } \rightarrow - \omega _ { \eta q } , \omega_{\chi k} \rightarrow \omega_{\chi k} , \omega_{\chi l} \rightarrow \omega_{\chi l} \right)\\
		\approx& -\frac{\partial}{\partial \omega}\Bigg\{\frac { 1 } { 4 ( 2 \pi ) ^ { 3 } } \int _ { M_\chi } ^ { \infty } d \omega_{\chi l} \int _ { M_\chi t _ { 1 } - } ^ { M_\chi t _ { 1 + } } d \omega_{\chi k}\big[\gamma _ { i } \left( - \omega + \omega_{\chi k} + \omega_{\chi l} , - \omega_{\chi k} , - \omega_{\chi l} \right)\\
		&\times \left( 1 + f \left( \omega_{\chi k} + \omega_{\chi l} - \omega \right) \right) f \left( \omega_{\chi k} \right) f \left( \omega_{\chi l} \right)\\
		& \quad - \gamma _ { i } \left( \omega - \omega_{\chi k} - \omega_{\chi l} , \omega_{\chi k} , \omega_{\chi l} \right) f \left( \omega_{\chi k} + \omega_{\chi l} - \omega \right) \left( 1 + f \left( \omega_{\chi k} \right) \right) \left( 1 + f \left( \omega_{\chi l} \right) \right)\big]+\left(\omega\to -\omega\right)\Bigg\}
	\end{aligned}
\end{equation}
where
\begin{equation}
	\left\{
	\begin{aligned}
		& t _ { 1 \pm } = ( u - 1 ) s \pm \sqrt { u ( u - 2 ) \left( s ^ { 2 } - 1 \right) } + \frac { \omega } { M_\chi } 
		\left[ 1 - u + ( 2 u - 1 ) s ^ { 2 } \pm s \sqrt { u ( u - 2 ) \left( s ^ { 2 } - 1 \right) \left( \frac { 2 u - 3 } { u - 2 } \right) } \right]\\
		& u = \frac { M_\eta ^ { 2 } } { 2 M_\chi^ { 2 } } , s = \frac { \omega_{\chi l} } {M_\chi}
	\end{aligned}
	\right.~.
\end{equation}

We shall presently compute the derivatives of $ I_s $ and $ I_c $ and setting $\omega=0$:
\begin{equation}\label{I_s}  
	\begin{aligned}
		 \left. \frac { \partial } { \partial \omega } I _ { s } \left[ \gamma _ { i } \right] \right|_ { \omega = 0 }
		= &2 \cdot \frac { 1 } { 4 ( 2 \pi ) ^ { 3 } } \int _ {M_\chi} ^ { \infty } d \omega_{\chi l}\int_{M_\chi \left[ ( u - 1 ) s - \sqrt { u ( u - 2 ) \left( s ^ { 2 } - 1 \right) } \right]}^{M_\chi \left[ ( u - 1 ) s + \sqrt { u ( u - 2 ) \left( s ^ { 2 } - 1 \right) } \right]}d \omega_{\chi k}\\
		& ~~\frac { \partial } { \partial \omega } \bigg[\gamma _ { i } \left( - \omega + \omega_{\chi k} + \omega_{\chi l} , - \omega_{\chi k} , - \omega_{\chi l} \right) \left( 1 + f \left( \omega_{\chi k} + \omega_{\chi l} - \omega \right) \right) f \left( \omega_{\chi k} \right) f \left( \omega_{\chi l} \right)\\
		& ~~- \gamma _ { i } \left( \omega - \omega_{\chi k} - \omega_{\chi l} , \omega_{\chi k} , \omega_{\chi l} \right) f \left( \omega_{\chi k} + \omega_{\chi l} - \omega \right) \left( 1 + f \left( \omega_{\chi k} \right) \right) \left( 1 + f \left( \omega_{\chi l} \right) \right)\bigg]_{\omega=0}\\
		+& 2 \cdot \frac { 1 } { 4 ( 2 \pi ) ^ { 3 } } \int _ {M_\chi} ^ { \infty } d \omega_{\chi l}\bigg\{M_\chi t_{1p}\Big[\gamma _ { i } \left( \omega_{\chi k} + \omega_{\chi l} , - \omega_{\chi k} , - \omega_{\chi l} \right) \left( 1 + f \left( \omega_{\chi k} + \omega_{\chi l} \right) \right) f \left( \omega_{\chi k} \right) f \left( \omega_{\chi l} \right)\\
		& - \gamma _ { i } \left( - \omega_{\chi k} - \omega_{\chi l} , \omega_{\chi k} , \omega_{\chi l} \right) f \left( \omega_{\chi k} + \omega_{\chi l} \right) \left( 1 + f \left( \omega_{\chi k} \right) \right) \left( 1 + f \left( \omega_{\chi l} \right) \right)\Big]_{\omega_{\chi k}=M_\chi t_{0p}}\\
		& - M_\chi t _ { 1 m }\Big[\gamma _ { i } \left( \omega_{\chi k} + \omega_{\chi l} , - \omega_{\chi k} , - \omega_{\chi l} \right) \left( 1 + f \left( \omega_{\chi k} + \omega_{\chi l} \right) \right) f \left( \omega_{\chi k} \right) f \left( \omega_{\chi l} \right)\\
		&- \gamma _ { i } \left( - \omega_{\chi k} - \omega_{\chi l} , \omega_{\chi k} , \omega_{\chi l} \right) f \left( \omega_{\chi k} + \omega_{\chi l} \right) \left( 1 + f \left( \omega_{\chi k} \right) \right) \left( 1 + f \left( \omega_{\chi l} \right) \right)\Big]_{\omega_{\chi k}=M_\chi t_{0m}}\bigg\}~.
	\end{aligned}
\end{equation}
\begin{equation}  \label{I_c}  
	\begin{aligned}
		- \left. \frac { \partial ^ { 2 } } { \partial \omega ^ { 2 } } I _ { c } \left[ \gamma _ { i } \right] \right|_ { \omega = 0 }
		=& - \frac { 1 } { 2 ( 2 \pi ) ^ { 3 } } \int _ { M_\chi } ^ { \infty } d \omega_{\chi l}\Bigg\{\int _ { M_\chi t _ { 0 m } } ^ { M_\chi t _ { 0 p } } d \omega_{\chi k} \frac { \partial ^ { 2 } } { \partial \omega ^ { 2 } }\Big[\gamma _ { i } \left( - \omega + \omega_{\chi k} + \omega_{\chi l} , - \omega_{\chi k} , - \omega_{\chi l} \right) \\
		&\times\left( 1 + f \left( \omega_{\chi k} + \omega_{\chi l} - \omega \right) \right) f \left( \omega_{\chi k} \right) f \left( \omega_{\chi l} \right)\\
		& - \gamma _ { i } \left( \omega - \omega_{\chi k} - \omega_{\chi l} , \omega_{\chi k} , \omega_{\chi l} \right) f \left( \omega_{\chi k} + \omega_{\chi l} - \omega \right) \left( 1 + f \left( \omega_{\chi k} \right) \right) \left( 1 + f \left( \omega_{\chi l} \right) \right)\Big]\bigg|_{\omega=0}\\
		& +M_\chi^2(t_{1p})^2\frac{\partial}{\partial \omega_{\chi k}}\Big[\gamma_i(\omega_{\chi k}+\omega_{\chi l},-\omega_{\chi k},-\omega_{\chi l})\left(1+f\left(\omega_{\chi k}+\omega_{\chi l}\right)\right)f\left(\omega_{\chi k}\right)f\left(\omega_{\chi l}\right)\\
		&-\gamma_i\left(-\omega_{\chi k}-\omega_{\chi l},\omega_{\chi k},\omega_{\chi l}\right)f\left(\omega_{\chi k}+\omega_{\chi l}\right)\left(1+f\left(\omega_{\chi k}\right)\right)\left(1+f\left(\omega_{\chi l}\right)\right)\Big]\bigg|_{\omega_{\chi k}\to M_\chi t_{0p}}\\
		&-M_\chi^2(t_{1m})^2\frac{\partial}{\partial \omega_{\chi k}}\Big[\gamma_i(\omega_{\chi k}+\omega_{\chi l},-\omega_{\chi k},-\omega_{\chi l})\left(1+f\left(\omega_{\chi k}+\omega_{\chi l}\right)\right)f\left(\omega_{\chi k}\right)f\left(\omega_{\chi l}\right)\\
		&-\gamma_i\left(-\omega_{\chi k}-\omega_{\chi l},\omega_{\chi k},\omega_{\chi l}\right)f\left(\omega_{\chi k}+\omega_{\chi l}\right)\left(1+f\left(\omega_{\chi k}\right)\right)\left(1+f\left(\omega_{\chi l}\right)\right)\Big]\bigg|_{\omega_{\chi k}\to M_\chi t_{0m}}\Bigg\},  
	\end{aligned}
\end{equation}
where  
\begin{equation}   
\left\{
	\begin{aligned}
		& u=\frac{M_\eta^2}{2M_\chi^2}, s=\frac{\omega_{\chi l}}{M_\chi}\\
		& t _ { 0 p } = ( u - 1 ) s + \sqrt { u ( u - 2 ) \left(  s  ^ { 2 } - 1 \right) }\\
		& t _ { 0 m } = ( u - 1 ) s - \sqrt { u ( u - 2 ) \left( s  ^ { 2 } - 1 \right) }\\
		& t _ { 1 p } = \frac { 1 } { M_\chi } \left[ 1 - u + ( 2 u - 1 ) s ^ { 2 } + s \sqrt { u ( u - 2 ) \left( s ^ { 2 } - 1 \right) } \left( \frac { 2 u - 3 } { u - 2 } \right) \right]\\
		& t _ { 1 m } = \frac { 1 } { M_\chi } \left[ 1 - u + ( 2 u - 1 ) s ^ { 2 } - s \sqrt { u ( u - 2 ) \left( s ^ { 2 } - 1 \right) } \left( \frac { 2 u - 3 } { u - 2 } \right) \right]\\
		& t _ { 1 + } = t _ { 0 p } + \omega t _ { 1 p } , \quad t _ { 1 - } = t _ { 0 m } + \omega t _ { 1 m }~.
	\end{aligned}\right.
\end{equation}

We  also  calculate the derivatives of  the Fourier transformations of, 
 $ t\sin(\omega_{\eta q}+\omega_{\chi k}+\omega_{\chi l}) $ and $ t^2\cos(\omega_{\eta q}+\omega_{\chi k}+\omega_{\chi l}) $  (which we denote by $ J_s $ and $ J_c $, respectively)  and then evaluate them at $ \omega=0 $:
\begin{equation}\label{J_s}
	\begin{aligned}
  \left.\frac { \partial } { \partial \omega } J _ { s } \left[ \gamma _ { i } \right]\right| _ { \omega = 0 } \approx 
		& - 2 \frac { 1 } { ( 2 \pi ) ^ { 4 } } \int _ { M_\chi } ^ { \infty } d \omega_{\chi k} d \omega_{\chi l} \int _ { - 1 } ^ { 1 } d w \frac { \sqrt { \omega_{\chi k} ^ { 2 } - M_\chi ^ { 2 } } \sqrt { \omega_{\chi l} ^ { 2 } - M_\chi ^ { 2 } }  \left( \omega _ { \eta q } + \omega_{\chi k} + \omega_{\chi l} \right) ^ 3} { \omega _ { \eta q } \left[\left( \omega _ { \eta q } + \omega_{\chi k} + \omega_{\chi l} \right) ^2+\left(\Gamma_\chi(\omega_{\chi k})/2+\Gamma_\chi(\omega_{\chi l})/2\right)^2\right]  ^3}\\
		&\times  \gamma _ { i } \left( \omega _ { \eta q } , \omega_{\chi k} , \omega_{\chi l} \right)\left( 1 + f (\omega_{\eta q}) \right) \left( 1 + f (\omega_{\chi k})\right) \left( 1 + f (\omega_{\chi l}) \right)\\
		& + \left( \omega _ { \eta q } \rightarrow - \omega _ { \eta q } , \omega_{\chi k} \rightarrow - \omega_{\chi k} , \omega_{\chi l} \rightarrow - \omega_{\chi l} \right)\\
		& + \left( \omega _ { \eta q } \rightarrow - \omega _ { \eta q } , \omega_{\chi k} \rightarrow - \omega_{\chi k} , \omega_{\chi l} \rightarrow \omega_{\chi l} \right) + \left( \omega _ { \eta q } \rightarrow \omega _ { \eta q } , \omega_{\chi k} \rightarrow \omega_{\chi k} , \omega_{\chi l} \rightarrow - \omega_{\chi l} \right)\\
		& + \left( \omega _ { \eta q } \rightarrow - \omega _ { \eta q } , \omega_{\chi k} \rightarrow \omega_{\chi k} , \omega_{\chi l} \rightarrow - \omega_{\chi l} \right) + \left( \omega _ { \eta q } \rightarrow \omega _ { \eta q } , \omega_{\chi k} \rightarrow - \omega_{\chi k} , \omega_{\chi l} \rightarrow \omega_{\chi l} \right)\\
		& + \left( \omega _ { \eta q } \rightarrow \omega _ { \eta q } , \omega_{\chi k} \rightarrow - \omega_{\chi k} , \omega_{\chi l} \rightarrow - \omega_{\chi l} \right) + \left( \omega _ { \eta q } \rightarrow - \omega _ { \eta q } , \omega_{\chi k} \rightarrow \omega_{\chi k} , \omega_{\chi l} \rightarrow \omega_{\chi l} \right)~,
	\end{aligned}
\end{equation}
\begin{equation}\label{J_c}
	\begin{aligned}
	 \left.\frac { \partial } { \partial \omega } J _ { c }  \left[ \gamma _ { i } \right] \right| _ { \omega = 0 }
		& \approx 6 \frac { 1 } { ( 2 \pi ) ^ { 4 } } \int _ { M_\chi } ^ { \infty } d \omega_{\chi k} d \omega_{\chi l} \int _ { - 1 } ^ { 1 } d w \frac { \sqrt { \omega_{\chi k} ^ { 2 } - M_\chi ^ { 2 } } \sqrt { \omega_{\chi l} ^ { 2 } - M_\chi ^ { 2 } }  \left( \omega _ { \eta q } + \omega_{\chi k} + \omega_{\chi l} \right) ^ 4} { \omega _ { \eta q } \left[\left( \omega _ { \eta q } + \omega_{\chi k} + \omega_{\chi l} \right) ^2+\left(\Gamma_\chi(\omega_{\chi k})/2+\Gamma_\chi(\omega_{\chi l})/2\right)^2\right]  ^4} \\
		&\times \gamma _ { i } \left( \omega _ { \eta q } , \omega_{\chi k} , \omega_{\chi l} \right)\left( 1 + f (\omega_{\eta q}) \right) \left( 1 + f (\omega_{\chi k})\right) \left( 1 + f (\omega_{\chi l}) \right)\\
		& + \left( \omega _ { \eta q } \rightarrow - \omega _ { \eta q } , \omega_{\chi k} \rightarrow - \omega_{\chi k} , \omega_{\chi l} \rightarrow - \omega_{\chi l} \right)\\
		& + \left( \omega _ { \eta q } \rightarrow - \omega _ { \eta q } , \omega_{\chi k} \rightarrow - \omega_{\chi k} , \omega_{\chi l} \rightarrow \omega_{\chi l} \right) + \left( \omega _ { \eta q } \rightarrow \omega _ { \eta q } , \omega_{\chi k} \rightarrow \omega_{\chi k} , \omega_{\chi l} \rightarrow - \omega_{\chi l} \right)\\
		& + \left( \omega _ { \eta q } \rightarrow - \omega _ { \eta q } , \omega_{\chi k} \rightarrow \omega_{\chi k} , \omega_{\chi l} \rightarrow - \omega_{\chi l} \right) + \left( \omega _ { \eta q } \rightarrow \omega _ { \eta q } , \omega_{\chi k} \rightarrow - \omega_{\chi k} , \omega_{\chi l} \rightarrow \omega_{\chi l} \right)\\
		& + \left( \omega _ { \eta q } \rightarrow \omega_{\eta q} , \omega_{\chi k} \rightarrow - \omega_{\chi k} , \omega_{\chi l} \rightarrow - \omega_{\chi l} \right) + \left( \omega_{\eta q} \rightarrow - \omega_{\eta q} , \omega_{\chi k} \rightarrow \omega_{\chi k} , \omega_{\chi l} \rightarrow \omega_{\chi l} \right)~.
	\end{aligned}
\end{equation}

The derivative of $ \kappa $ with respect to $ \omega $ can thus be expressed as,
\begin{equation}\label{self_imaginary1}
	\begin{aligned}
		& -i\left. \frac { \partial } { \partial \omega } \kappa ( t_1,\omega ) \right| _ { \omega = 0 }\\
		=&\frac{h^2}{8}\Bigg\{ \left( 1 - 9 H t _ { 1 } \right) \left. \frac { \partial } { \partial \omega } I _ { s }  \left[ \gamma _ { 0 } \right] \right| _ { \omega = 0 } + H t _ { 1 }\left.  \frac { \partial } { \partial \omega } I _ { s } \left[ \gamma _ { 1 } \right] \right| _ { \omega = 0 } - H t _ { 1 }\left.  \frac { \partial } { \partial \omega } I _ { s } \left[ \gamma _ { 2 } \right] \right| _ { \omega = 0 } - 2 H t _ { 1 }\left.  \frac { \partial ^ { 2 } } { \partial \omega ^ { 2 } } I _ { c } \left[ \gamma _ { 3 } \right] \right|_ { \omega = 0 }\\
		& -\frac { 9 } { 2 } H\left.  \frac { \partial } { \partial \omega } J _ { s } \left[ \gamma _ { 0 } \right] \right| _ { \omega = 0 } - \frac { 1 } { 2 } H \left. \frac { \partial } { \partial \omega } J _ { s } \left[ \gamma _ { 1 } \right] \right|_ { \omega = 0 } + H \left. \frac { \partial } { \partial \omega } J _ { s } \left[ \gamma _ { 2 } \right] \right| _ { \omega = 0 } - H\left.  \frac { \partial } { \partial \omega } J _ { c } \left[ \gamma _ { 3 } \right] \right|_ { \omega = 0 }\Bigg\}.
	\end{aligned}
\end{equation}

\addcontentsline{toc}{section}{References}

\bibliographystyle{JHEP}
\providecommand{\href}[2]{#2}
\begingroup\raggedright

\endgroup
	
\end{document}